\documentclass[preprint,12pt]{elsarticle}

\usepackage{amssymb}
\usepackage{amsmath}
\journal{Theoretical and Applied Fracture Mechanics}

\begin{document}

\begin{frontmatter}



\title{The role of weak interfaces in the tensile deformation and fracture of particle-filled polymers studied by phase-field model}


\author[BUW,LUH]{Xu Chen}

\author[BUW]{Ya Duan}

\author[TJU,LUH]{Xiaoying Zhuang}

\author[BUW,FDU]{Timon Rabczuk\corref{cor}}
\ead{timon.rabczuk@uni-weimar.de}

\cortext[cor]{Corresponding author}

\affiliation[BUW]{
  organization={Institute of Structural Mechanics, Bauhaus-Universität Weimar},
  city={Weimar},
  postcode={99423}, 
  country={Germany}
}

\affiliation[LUH]{
  organization={Institute of Photonics, Department of Mathematics and Physics, Leibniz University Hannover},
  city={Hannover},
  postcode={30167}, 
  country={Germany}
}

\affiliation[TJU]{
  organization={Department of Geotechnical Engineering, College of Civil Engineering, Tongji University},
  city={Shanghai},
  postcode={200092}, 
  country={China}
}

\affiliation[FDU]{
  organization={School of Intelligent Robotics and Advanced Manufacturing, Fudan University},
  city={Shanghai},
  postcode={200433}, 
  country={China}
}

%\affiliation{organization={},%Department and Organization

%% Abstract
\begin{abstract}
Weak particle-matrix interfaces play a critical role in the tensile fracture of particle-filled polymer composites, but how they govern progressive debonding, fracture localization, and the resulting changes in macroscopic mechanical properties remains insufficiently understood. In this study, a cohesive-zone phase-field model incorporating a hyperelastic polymer matrix and a smeared interface is employed to investigate the coupled evolution of interfacial debonding and matrix fracture in particle-filled polymer composites. The model is calibrated against and compared with uniaxial tensile responses of particle-filled polyurethane composites and then used to study how interfacial strength, interfacial fracture energy, and matrix fracture properties affect the macroscopic stress-strain response and damage evolution. The results show that weak interfaces can induce an intermediate softening regime in the stress-strain response, characterized by a reduced effective tangent stiffness and associated with distributed interfacial damage. Interfacial strength mainly controls the initiation of debonding, whereas interfacial fracture energy affects whether debonding can develop progressively in a distributed manner or rapidly localizes into a dominant crack band. Comparisons with well-bonded reference systems further demonstrate that weak interfaces may reduce the maximum stress but increase the strain at break by promoting distributed debonding around particles and delaying the formation of a dominant crack band. These findings clarify the dual role of weak interfaces and provide a mechanistic understanding of interface-controlled tensile failure in particle-filled polymer composites.
\end{abstract}



%% Keywords

\end{frontmatter}


% \linenumbers
\section{Introduction}

Particle-filled polymer composites consist of a continuous polymer matrix containing rigid particles and are widely used in engineering products such as tires, coatings, adhesives, and electrical components. Their mechanical properties, including the elastic modulus, tensile strength, toughness, and strain at break, are affected by the introduction of particles and thus depend strongly on the particle size and morphology, filler loading, and particle-matrix interfacial adhesion\cite{Fu2008,Bommegowda2021}. Compared with the pure polymer, the introduction of particles creates significant microstructural and mechanical heterogeneity because of the mismatch in stiffness and deformation capacity between particles and matrix. This heterogeneity affects local deformation and damage processes, which in turn influence the macroscopic mechanical response of the composite\cite{Tao2013}. Consequently, improving these composites requires understanding how microstructure, local deformation, and damage evolution affect their macroscopic response and fracture.

In particle-filled polymer composites, the particle-matrix interface is a key microstructural feature governing load transfer and damage evolution\cite{Kashfipour2018}. Because particles and matrix often differ in surface chemistry and deformation capacity, the interface can become a weak region where damage initiates under tensile loading, unless suitable surface treatments or coupling agents are used\cite{LeGulluche2023}. Experiments have also shown that damage often starts as particle-matrix debonding, or nearby matrix cavitation and cracking\cite{Gent1984,Poulain2017}. Once initiated, damage may first extend along the particle-matrix interface and then spread into the surrounding matrix. With continued loading, neighboring damaged zones may coalesce into a major macroscopic fracture path, ultimately leading to failure\cite{Toulemonde2016}. Consequently, the fracture properties of the interface strongly affect damage evolution and the mechanical properties of the composite.

Although interfacial debonding and crack growth in polymer composites have been extensively investigated, the effect of interfacial properties on macroscopic tensile deformation and fracture cannot be described by a simple monotonic relation\cite{Fu2008,Kundie2018,Kun2021}. Weak interfaces generally debond at low loads, weakening stress transfer and usually reducing stiffness, tensile strength, or elongation at break\cite{Zebarjad2004,Fan2015}. Accordingly, filler surface modification and coupling agents are widely used to strengthen particle-matrix adhesion\cite{Rong2006,Ippolito2020,Bi2022}. However, weak adhesion can also improve toughness or ductility in some cases. Thio et al. varied the adhesion between glass particles and a polypropylene matrix and showed that weakly adhering surfaces increased the macroscopic toughness of the composite\cite{Thio2004}. Aliotta et al. found that enhanced ductility required interfacial debonding to be accompanied by stable and controlled void growth\cite{Aliotta2019}. These observations indicate that the macroscopic failure response is not determined by interfacial debonding alone, but also by the subsequent evolution of damage\cite{Hsieh2010,Quaresimin2016}, which may either promote stable deformation and energy dissipation, or progress toward damage coalescence and fracture localization. However, it remains insufficiently understood how interfacial fracture properties regulate this mesoscale damage process, and why weak interfaces can lead to different or opposite changes in macroscopic mechanical properties for polymer composite.

Beyond their effects on strength and strain at break, weak interfaces can also alter the macroscopic stress--strain response of particle-filled polymers under uniaxial tension. An intermediate softening regime is usually associated with progressive and spatially distributed debonding around multiple filler particles. In this regime, as the interfacial load-transfer capacity decreases, the effective tangent stiffness of the composite is reduced, and the macroscopic response deviates from that of the undamaged composite\cite{Vollenberg1988,Asp1997,Renner2005}. With continued loading, damage coalescence and the formation of a dominant fracture path leads to an abrupt stress drop associated with macroscopic fracture\cite{Cheng2007}. Recent advances in \textit{in situ} imaging have enabled more direct characterization of internal damage initiation and evolution, including interfacial debonding, cavitation, and crack development, thereby providing further insight into the microscopic mechanisms underlying the observed tensile response\cite{Gilormini2017,Lai2024}. Despite these advances, the respective roles of interfacial and matrix fracture properties in governing the different stages of damage evolution and the macroscopic stress--strain response remain insufficiently understood.

Experimental studies have widely investigated the effects of filler content, particle size, and interfacial conditions on the mechanical response and damage of particle-filled polymers. However, material and microstructural variables are often coupled and cannot be independently varied in experiments\cite{Aliotta2019,Liu2015}, making it difficult to isolate the respective contributions of the interface and matrix. Modeling provide a complementary approach by allowing individual parameters to be studied independently. While analytical mean-field schemes efficiently predict effective properties, they generally cannot resolve damage localization and crack-path evolution\cite{Brassart2009,Firooz2021}. In contrast, mesoscale representative volume elements (RVE) simulations explicitly represent particle arrangements and local damage evolution\cite{deFrancqueville2020}. Previous finite-deformation RVE studies have shown that interfacial strength mainly affects damage initiation and peak stress, whereas interfacial fracture energy influences subsequent damage evolution and failure\cite{Moraleda2009,Toulemonde2016}. However, many such models focus on cohesive debonding along predefined particle-matrix interface, while matrix fracture is omitted or treated through a separate failure criterion rather than as an independently evolving fracture process.

Therefore, a more general mesoscale framework is needed to capture the coupled evolution of interfacial debonding and matrix fracture without prescribing potential fracture paths in advance. Continuum variational approaches provide an alternative route by representing evolving fracture through continuous fields rather than explicitly inserted discontinuities. These include phase-field fracture formulations\cite{Ambati2015} and more recent variational damage formulations\cite{Ren2024,Duan2026,Duan20262}. In the phase-field method, a sharp crack is regularized into a smeared crack band of finite width, allowing crack initiation, propagation, branching, and coalescence to be simulated naturally. The development of the phase-field approach was pioneered by the works of Francfort and Marigo\cite{Francfort1998}, Bourdin et al.\cite{Bourdin2000}, and Miehe et al.\cite{Miehe2010}. However, for heterogeneous materials, fracture along material interfaces requires additional treatment. Two main approaches are commonly adopted. In hybrid phase-field/cohesive-element formulations, bulk fracture is described by a phase-field model and interfacial debonding is modeled by cohesive-zone elements explicitly inserted along material interfaces\cite{Paggi2017,Zhang2020,Najmeddine2025}, retaining a sharp interface representation and a direct traction-separation description. In diffuse-interface phase-field formulations, the sharp material interface is regularized as a smeared interfacial zone and incorporated into the phase-field framework\cite{HansenDrr2019,HansenDrr2020}, allowing a common crack phase field to describe both interfacial debonding and bulk fracture\cite{Nguyen2016,Zhang2019,HansenDrr2020,Chen2025}. Compared with hybrid formulations, this unified treatment avoids the need to establish coupling relations between separate bulk and interface fracture models and simplifies the numerical treatment of their interaction, allowing crack nucleation, interfacial debonding, crack penetration, branching, and coalescence to evolve more naturally within a common framework\cite{Chen2025}. When applied to polymer composites, the fracture framework also needs to account for the large-deformation behavior of the polymer matrix. Phase-field fracture models for elastomers and other polymeric solids have been extended to finite-deformation settings through hyperelastic formulations\cite{Miehe2014,Mandal2020}, rate-dependent viscoelastic formulations\cite{Loew2019,Damma2023}, and polymer molecular-informed damage formulations\cite{Talamini2018}, demonstrating the applicability of phase-field approaches to polymer fracture under different constitutive descriptions and damage-driving mechanisms. A further issue is that conventional phase-field models intrinsically couple the apparent material strength to the fracture energy and regularization length scale, requiring the regularization length scale to be selected according to the target material strength\cite{Tann2018}. To overcome this limitation, cohesive phase-field formulations explicitly introduce fracture strength as an independent material parameter and reproduce prescribed cohesive traction-separation laws, thereby reducing the sensitivity of the simulation results to the regularization length scale\cite{Wu2017,Wu2018,Geelen2019}. Together, these developments provide the basis for studying the coupled evolution of interfacial debonding and matrix fracture in polymer composites.

To clarify how weak particle-matrix interfaces influence the macroscopic tensile properties and stress-strain response of particle-filled polymer composites, this study employs a CZM phase-field model to systematically investigate deformation and damage evolution in RVEs. The formulation combines a hyperelastic constitutive model with a smeared representation of the sharp particle-matrix interface, allowing interfacial debonding and matrix fracture to be described by a common crack phase field. The model is first verified using a one-dimensional bar problem and a single-fiber benchmark under transverse tension, followed by sensitivity analyses of the RVE simulations. The model is then calibrated and compared with experimental uniaxial tensile stress-strain responses of particle-filled polyurethane composites with different filler volume fractions. Studies are subsequently carried out to illustrate how interfacial strength, interfacial fracture energy, and matrix fracture properties govern damage initiation and evolution, resulting multi-stage stress-strain response. Finally, by comparing weakly bonded composites with well-bonded reference systems, it quantifies the influence of interfacial weakening on the macroscopic mechanical response, particularly the strain at break, and identifies the conditions under which a weak interface increases or decreases the strain at break, providing mechanistic insight into the contrasting effects of weak interfaces in different systems.

\section{Model Formulation for Bulk and Interfacial Damage}

\begin{figure}[htpb]
    \centering
    \includegraphics[width=1.0\linewidth]{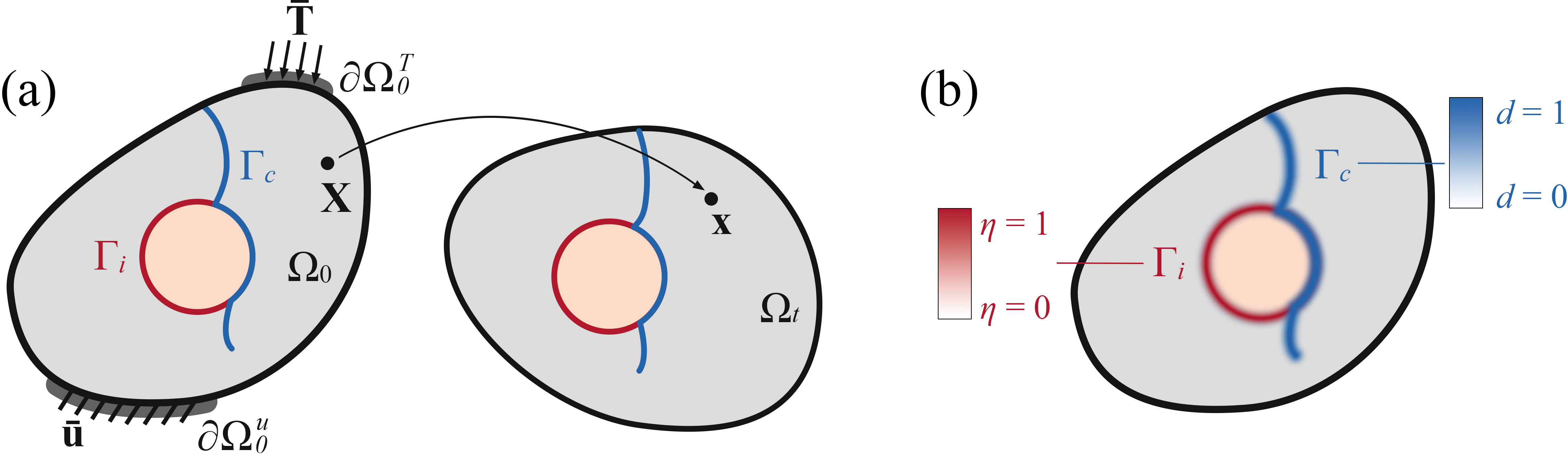}
    \caption{(a) Reference configuration $\Omega_0$ of the composite containing an inclusion surrounded by an internal interface $\Gamma_i$ and a crack surface $\Gamma_c$. A material point with position $\mathbf{X}$ in $\Omega_0$ is mapped to its current position $\mathbf{x}$ in the deformed configuration $\Omega_t$. (b) Regularization of the sharp interface and crack surface by the scalar phase fields $\eta$ and $d$, respectively.}
    \label{fig:PFM_model}
\end{figure}

Consider a particle-filled polymer composite represented by a domain $\Omega_0$ containing an inclusion surrounded by an internal interface $\Gamma_i$. In addition, a crack surface $\Gamma_c$ may be present in the matrix and evolve under external tensile loading. Displacement boundary conditions $\bar{\mathbf{u}}$ are applied on $\partial\Omega^u_0$, and tractions $\bar{\mathbf{T}}$ are applied on $\partial\Omega^T_0$. The vector $\mathbf{X}$ denotes the position of an arbitrary point in $\Omega_0$, and its current position at time $t$ in the deformed configuration $\Omega_t$ is denoted by $\mathbf{x}$. The geometric configuration and boundary conditions are illustrated in Fig.~\ref{fig:PFM_model}(a). Explicitly resolving the sharp interface and tracking the crack path can be numerically challenging, particularly when interfacial debonding interacts with matrix cracking. To avoid the explicit representation of these sharp surfaces, a phase-field framework similar to a previously proposed formulation is adopted\cite{Zhen2024}, in which the material interface $\Gamma_i$ is represented by a stationary smeared interface field $\eta$, which is determined by the initial composite geometry and kept fixed during mechanical loading, while the evolving crack surface $\Gamma_c$ is represented by the crack phase field $d$, as sketched in Fig.~\ref{fig:PFM_model}(b). 

\subsection{Phase-field Formulation for Fracture of Hyperelastic Material}

The phase-field description of fracture in a hyperelastic solid starts from the total potential energy $\Psi$, which can be written as the sum of the elastic strain energy $\Psi_{\rm ela}$, the fracture energy $\Psi_{\rm frac}$, and the external-load potential $\Psi_{\rm ext}$.
\begin{equation}
    \label{eq:total_energy}
    \Psi = \Psi_{\rm ela} + \Psi_{\rm frac} + \Psi_{\rm ext}
\end{equation}

For the fracture contribution, the scalar phase-field variable $d$ is used to describe material damage, where $d = 0$ corresponds to the undamaged state and $d = 1$ corresponds to the fully damaged state. The crack surface is regularized by the crack surface density function $\gamma(d,\nabla d)$ \cite{Miehe2010}, so that the fracture energy can be written as
\begin{equation}
    \Psi_{\rm frac} = \int_{\Gamma_c} G_c\ {\rm d}A = \int_{\Omega_0} G_c \gamma(d, \nabla d) \ {\rm d}V
\end{equation}
where $G_c$ denotes the critical energy release rate of the material. 

Within the phase-field framework, fracture energy dissipation and the associated material softening behavior are largely governed by the choice of the crack surface density function $\gamma$. To decouple the material strength and softening response from the length scale $l_c$, the phase-field framework inspired by the Cohesive-Zone Model (CZM) developed by Wu et al.\cite{Wu2017,Wu2018} is adopted. In this approach, the crack surface density function is defined as
\begin{equation}
    \gamma(d, \nabla d) = \frac{1}{c_0}\left( \frac{\alpha(d)}{l_c} + l_c|\nabla d|^2 \right)
\end{equation}
where $l_c$ is the characteristic length controlling the width of the regularized crack, $c_0$ is the scaling parameter, and $\alpha(d)$ is the geometric crack function. In this work, $\alpha(d) = 2d - d^2$ is used, which gives $c_0 = \pi$.

Since the present study focuses on the overall deformation and fracture process rather than time-dependent effects, the polymer matrix is modeled as a hyperelastic solid in a finite-deformation setting. The deformation mapping $\mathbf{x} = \boldsymbol{\varphi}(\mathbf{X}, t)$ maps the reference configuration $\Omega_0$ to the current configuration $\Omega_t$. The associated deformation gradient is $\mathbf{F} = \partial\mathbf{x}/\partial\mathbf{X}$, the right Cauchy-Green deformation tensor is $\mathbf{C} = \mathbf{F}^\top\mathbf{F}$, and the Jacobian is $J = \det \mathbf{F}$. A compressible Neo-Hookean model is adopted as a simple yet representative hyperelastic law\cite{Zienkiewicz2000}, with the undamaged strain energy density given by
\begin{equation}
    \label{eq:NH_model}
    \psi_0(\mathbf{F}) = \frac{\mu}{2} ({\rm tr}(\mathbf{C}) - 3 - 2 \log(J)) + \frac{\lambda}{2} (\log J) ^2
\end{equation}
where ${\rm tr}(\cdot)$ denotes the trace operator, and $\mu$ and $\lambda$ denote the Lam\'e parameters. 

During fracture evolution, stiffness degradation due to cracking is accounted for by multiplying a scalar degradation function $\omega(d)$ to the undamaged Neo-Hookean energy density $\psi_0(\mathbf{F})$, and the elastic part of the total energy is written as
\begin{equation}
    \Psi_{\rm ela} = \int_{\Omega_0} \omega(d)\psi_0(\mathbf{F})\ {\rm d}V
\end{equation}
where $\omega(d)$ is a degradation function that depends on the specific phase-field formulation. Within the CZM-inspired phase-field framework \cite{Wu2017,Wu2018}, it is defined as
\begin{equation}
    \omega = \frac{(1-d)^p}{(1-d)^p + (a_1d + a_1a_2d^2 + a_1a_2a_3d^3)}
\end{equation}
where the coefficients $p, a_1, a_2, a_3$ affect the softening behavior once the crack is initiated. In this work, the simple linear softening law is adopted \cite{Wu2017,Wu2018}, with $p = 2$, $a_2 = -0.5$, $a_3 = 0$, and
\begin{equation}
    a_1 = \frac{4l_{ch}}{\pi l_c}
\end{equation}
where $l_{ch}$ is the Griffith's or Irwin's internal length. For a linear elastic material with Young's modulus $E_0$, tensile strength $f_t$, and fracture energy $G_c$, the characteristic length scale is defined as $l_{ch}=E_0 G_c/f_t^2$. For a hyperelastic material with a nonlinear stress-strain response, this expression cannot be used directly. Following Mandal et al.\cite{Mandal2020}, an equivalent tensile strength $f_{te}$ is introduced to retain the same CZM phase field parameterization. It is defined by equating the elastic energy density of an equivalent linear material to the undamaged hyperelastic strain-energy density at damage initiation \cite{Mandal2020},
\begin{equation}
    f_{te}=\sqrt{2E_0\psi_0(\sigma^*)}
\end{equation}
where $\sigma^*$ is the critical Cauchy stress under uniaxial tension, and $\psi_0(\sigma^*)$ denotes the undamaged strain energy density of the hyperelastic material at the corresponding critical deformation. Thus, the Irwin's internal length becomes $l_{ch}=E_0G_c/f_{te}^2$. Given the critical tensile stress $\sigma^*$ and the undamaged constitutive relation $\psi_0$, the equivalent tensile strength $f_{te}$ can be calculated, and the coefficient $a_1$ in the degradation function can then be determined.

The potential of the external loads is written as
\begin{equation}
    \Psi_{\rm ext} = -\int_{\Omega_0} \mathbf{f}\cdot\mathbf{u}\ {\rm d}V - \int_{\partial \Omega^T_0} \bar{\mathbf{T}}\cdot\mathbf{u}\ {\rm d}A
\end{equation}
where $\mathbf{f}$ is the body force per unit reference volume, $\mathbf{u}$ is the displacement, and $\mathbf{\bar{T}}$ denotes the nominal traction vector on the Neumann boundary $\partial \Omega_0^T$.

For the above baseline energetic formulation, taking the first variation of Eq.\ref{eq:total_energy} with respect to $\mathbf{u}$ and $d$ yields
\begin{equation}
    \begin{split}
    \delta\Psi = &
    \int_{\Omega_0} \left(\omega(d) \mathbf{P}:\nabla_0 \delta\mathbf{u}- \mathbf{f}\cdot \delta\mathbf{u}\right) \ {\rm d}V 
    -\int_{\partial\Omega_0^T} \bar{\mathbf{T}} \cdot\delta\mathbf{u} \ {\rm d}A \\
    & + \int_{\Omega_0} \left[ \omega'(d)\psi_0(\mathbf{F})\delta d + \frac{G_c}{c_0}\left( \frac{\alpha'(d)}{l_c} \delta d + 2l_c \nabla_0d \cdot\nabla_0\delta d \right) \right] \ {\rm d} V 
    \end{split}
\end{equation}
where $\mathbf{P} = \partial\psi_0/\partial\mathbf{F}$ denotes the first Piola-Kirchhoff stress tensor, and $\nabla_0(\cdot)$ denotes the gradient with respect to the reference coordinates $\mathbf{X}$.

To prevent crack growth under compressive loading, the strain-energy density is split into tensile and compressive parts, denoted by $\psi^+$ and $\psi^-$, respectively. Only fracture driven by the tensile contribution is considered, so the total energy $\psi_0$ in the degradation term is replaced by $\psi^+$. For hyperelastic materials described by the Neo-Hookean constitutive model, we adopt a tension-compression split similar to those used in previous phase-field studies of hyperelastic-solid fracture\cite{Tang2019,Mandal2020}, which can effectively capture crack paths. The split is written as
\begin{equation}
    \begin{aligned}
        \psi^+ = \frac{\mu}{2} \sum_{i = 1}^3 ((\Lambda_i^+)^2 - 1 - & 2\log\Lambda_i^+) + \frac{\lambda}{2} (\log J^+) ^2  \\
        \psi^- = \frac{\mu}{2} \sum_{i = 1}^3 ((\Lambda_i^-)^2 - 1 - & 2\log\Lambda_i^-) + \frac{\lambda}{2} (\log J^-) ^2 
    \end{aligned}
\end{equation}
where $\Lambda_i$ denote the principal stretches, i.e., the square roots of the eigenvalues of the right Cauchy-Green tensor $\mathbf{C}$. The superscripts $(\cdot)^+$ and $(\cdot)^-$ denote the tensile and compressive contributions, respectively, and are defined as $(\cdot)^+ = \max((\cdot), 1)$ and $(\cdot)^- = \min((\cdot), 1)$.

Based on this tension-compression split, a hybrid formulation is adopted \cite{Ambati2015}. In this formulation, the tensile part $\psi^+$ is used only in the phase-field evolution equation, while the mechanical equilibrium equation uses the original unsplit hyperelastic response. This prevents the compressive part of the strain energy from leading to fracture evolution and reduces the complexity caused by the tension-compression split. To enforce damage irreversibility, $\psi^+$ is replaced by the local history field of the maximum tensile energy density $\max_{\tau\in[0,t]} \psi^+(\mathbf{F})$, which ensures $\dot{d}\ge0$. In the present model, damage should not evolve as long as the tensile energy density remains below the damage initiation threshold, so the phase-field variable should remain at $d = 0$ in this regime. This condition is imposed by introducing an energy threshold, so that small values of the tensile energy density are replaced by the constant value $\psi_0(\sigma^*)$, which is defined as the strain energy density corresponding to the critical failure stress. Overall, the effective energy density entering the phase-field equation is written as $H^+ = \max[\max_{\tau\in[0,t]} \psi^+(\mathbf{F}), \psi_0(\sigma^*)]$. With this tension-compression treatment, history field definition, and hybrid coupling strategy, the governing equations used in the present study are written in strong form as
\begin{equation}
\label{eq:strong_form}
        \left\{
    \begin{aligned}
    &\nabla_0 \cdot [\omega(d) \mathbf{P}] + \mathbf{f}=0 
        &&\text{in } \Omega_0 \\
    &\omega'(d)H^+ + \frac{G_c}{c_0}\left( \frac{\alpha'(d)}{l_c} - 2l_c\Delta_0 d \right) = 0 
        &&\text{in } \Omega_0 \\
    &\omega(d)\mathbf{P}\cdot\mathbf{N} = \bar{\mathbf{T}}
        &&\text{at } \partial\Omega^T_0 \\
    & \nabla_0 d\cdot\mathbf{N} = 0 
        &&\text{at } \partial\Omega_0 \\
    & \mathbf{u} = \bar{\mathbf{u}}
        &&\text{at } \partial\Omega^u_0
    \end{aligned}
    \right.
\end{equation}

\subsection{Smeared Interface Model for Internal Interfaces}

To represent the smeared interface, a stationary interface field $\eta(\mathbf{X})$ is constructed from the initial composite geometry, which takes values close to 1 within a narrow band around the interface and close to 0 in the bulk. This interface field $\eta$ only modulates material parameters such as $G_c$ and $\sigma^*$, whereas interfacial debonding itself is still described by the crack phase field $d$. Similar to the regularization of sharp crack path, the sharp interface is regularized by defining $\eta$ through the following strong form \cite{Zhang2019}
\begin{equation}
    \left\{
    \begin{aligned}
         &\eta - l_i^2 \Delta_0\eta = 0  && \text{in }\Omega_0  \\
         &\eta = 1 && \text{on interface}  \\
         &\nabla_0\eta\cdot \mathbf{N} = 0 && \text{at }\partial\Omega_0
    \end{aligned}
    \right.
\end{equation}
where $l_i$ is the characteristic length scale of the interface field. The length scale $l_i$ controls the width of the smeared interface zone and may influence the apparent interfacial toughness and crack-interface interactions in phase-field models\cite{HansenDrr2020}. Once the composite configuration is determined, this boundary-value problem is solved to obtain the stationary distribution of $\eta$, which is then kept fixed during the subsequent solution for the crack phase field and displacement field.

As $\eta$ decreases continuously from 1 at the interface toward 0 in the bulk, a general fracture-related parameter $q$ is interpolated between its interfacial value $q_i$ and matrix value $q_m$ as
\begin{equation}
    q(\eta) = [1 - (1-\eta)^2] q_i + (1-\eta)^2 q_m
\end{equation}
where $q_i$ and $q_m$ denote the interfacial and matrix values of the corresponding fracture-related parameter, respectively \cite{Zhang2019}. In this work, this interpolation is applied to the critical energy release rate $G_c$ and the tensile strength $\sigma^*$. The matrix and interfacial fracture energies are denoted by $G_m$ and $G_i$, respectively, and the corresponding tensile strengths by $\sigma^*_m$ and $\sigma^*_i$. Although the particle-matrix interfaces are locally subjected to combined normal and tangential loading, the present phase-field formulation does not explicitly distinguish between Mode I and Mode II fracture, and the interfacial strength and fracture energy are therefore treated as effective, mode-independent parameters.

For the interfacial critical energy release rate, the phase-field regularization should recover the same energy dissipation as the corresponding sharp-interface model when the crack propagates along the interface. To achieve this consistency, an effective critical energy release rate $\tilde{G}_i$ is introduced in place of $G_i$ in the interpolation of $G_c(\eta)$, so that $G_c$ becomes
\begin{equation}
    G_c(\eta) = [1 - (1-\eta)^2] \tilde{G}_i + (1-\eta)^2 G_m
\end{equation}
where $\tilde{G}_i$ should be calibrated so that the total dissipated energy equals $G_i$. To this end, a one-dimensional model containing the interface is considered to calculate $\tilde{G}_i$. For a fully developed crack on interface, the fracture energy obtained from the smeared-interface phase-field model should be equal to $G_i$, leading to
\begin{equation} 
\label{eq:effective_Gi}
\begin{split}
    G_i &= \int_{-b}^b G_c(\eta) \frac{1}{c_0}\left( \frac{\alpha(d)}{l_c} + l_c |\nabla d|^2 \right) \ {\rm d}x \\
        &= \int_{-b}^b \left[\left(1 - (1-\eta)^2\right)\tilde{G}_i + (1-\eta)^2 G_m \right] \frac{1}{c_0}\left( \frac{\alpha(d)}{l_c} + l_c |\nabla d|^2 \right) \ {\rm d}x
\end{split}
\end{equation}
where $b$ denotes the half-width of the diffuse crack band in this 1-d model. For the present PF-CZM model with linear softening law, the 1-d crack phase-field and interface field profiles are given by the analytical expressions\cite{Wu2017,Zhang2019}
\begin{equation} 
\label{eq:analytical_1d_field}
\begin{gathered}
    d(x) = 1 - \sin\left(\frac{|x|}{l_c}\right) \\
    \eta(x) = \exp(-|x|/l_i)
\end{gathered}
\end{equation}
with the half-width given by $b = \pi l_c/2$. Substituting these expressions into Eq.~\ref{eq:effective_Gi}, the effective value $\tilde{G}_i$ can be determined for given $G_i$ and $G_m$, ensuring that the smeared interface recovers the target energy dissipation associated with interfacial fracture.

In summary, the stationary interface field $\eta$ is first obtained from the initial composite geometry and then kept fixed during mechanical loading. The spatially varying fracture parameters $G_c(\eta)$ and $\sigma^*(\eta)$ are then calculated and incorporated into the governing equations for the displacement field $\mathbf{u}$ and the crack phase field $d$. With the present model, both interfacial debonding and matrix cracking are described by a common evolving crack phase field $d$, allowing cracks to initiate and propagate freely along the interface or into the matrix without prescribing an additional transition criterion. This provides a basis for investigating how interfacial fracture properties affect damage evolution and the macroscopic tensile response of particle-filled polymers.

\section{Numerical Implementation}

\subsection{Finite Element Discretization}

To solve the coupled phase-field problem numerically, the standard finite element method (FEM) is employed. The spatial discretization is carried out on a common mesh, and the displacement field, crack phase field, and interface field are interpolated within each element using the corresponding nodal values as
\begin{equation}
    \mathbf{u}(\mathbf{X}) = \mathbf{N}^u(\mathbf{X}) \mathbf{u}^e,\quad d(\mathbf{X}) = \mathbf{N}^d(\mathbf{X}) \mathbf{d}^e, \quad \eta (\mathbf{X}) = \mathbf{N}^\eta(\mathbf{X}) \boldsymbol{\eta}^e
\end{equation}
where $\mathbf{N}^q$ is the matrix of shape functions for field $q$ ($q \in \{u, d, \eta\}$). $\mathbf{u}^e$, $\mathbf{d}^e$, and $\boldsymbol{\eta}^e$ are the vectors of nodal values of the corresponding fields within the element. In this work, all two-dimensional simulations are performed under plane-strain conditions with a unit out-of-plane thickness of $1~\mathrm{mm}$, and the fields are discretized using meshes composed of linear 3-node triangular elements and 4-node bilinear quadrilateral elements. Numerical integration is performed using standard Gauss quadrature.

During the initialization step, the stationary interface field is obtained firstly from the weak form
\begin{equation}
    \int_{\Omega_0} \left[ \eta \delta \eta + l_i^2 \nabla_0 \eta \cdot \nabla_0 \delta \eta \right] \ {\rm d}V= 0
\end{equation}
Then $\eta$ is then fixed and the coupled weak forms for $\mathbf u$ and $d$ are
\begin{equation}
\begin{aligned}
    &\int_{\Omega_0} \omega(d) \mathbf{P}:\nabla_0 \delta\mathbf{u} \ {\rm d}V = \int_{\Omega_0} \mathbf{f}\cdot\delta\mathbf{u} \ {\rm d}V + \int_{\partial\Omega^T_0} \bar{\mathbf{T}}\cdot\delta\mathbf{u} \ {\rm d}A \\
    &\int_{\Omega_0} \left[ \omega'(d) H^+ \delta d + \frac{G_c(\eta)}{c_0}\left( \frac{2 - 2d}{l_c} \delta d + 2l_c \nabla_0d \cdot\nabla_0\delta d \right) \right] \ {\rm d} V = 0 \\
\end{aligned}
\end{equation}
With the Neo-Hookean constitutive model defined in Eq.\ref{eq:NH_model}, the first Piola-Kirchhoff stress can be derived as
\begin{equation}
    \mathbf{P} = \mu(\mathbf{F} - \mathbf{F}^{-\top}) + \lambda \log J\, \mathbf{F}^{-\top}
\end{equation}
Using the finite element interpolations, the residual vectors for the displacement field, crack phase field, and interface field are given by
\begin{equation}
\begin{aligned}
    &\mathbf{R}_u = -\int_{\Omega_e} \omega(d) (\mathbf{G}^u)^\top \{\mathbf{P}\} \ {\rm d}V + \int_{\Omega_e} (\mathbf{N}^u)^\top \mathbf{f} \ {\rm d}V + \int_{\partial\Omega_e^T} (\mathbf{N}^u)^\top \bar{\mathbf{T}} \ {\rm d} A \\
    &\mathbf{R}_d = -\int_{\Omega_e} \frac{G_c(\eta)}{c_0} \left[  \frac{2- 2d}{l_c}  ({\mathbf{N}^d})^\top + 2 l_c(\mathbf{B}^d)^\top \mathbf{B}^d \mathbf{d}^e \right] \ {\rm d}V 
    - \int_{\Omega_e} (\mathbf{N}^d)^\top \omega'(d) H^+ \ {\rm d}V \\
    &\mathbf{R}_\eta = - \int_{\Omega_e} \left[(\mathbf{N}^\eta)^\top \mathbf{N}^\eta  + l_i^2 (\mathbf{B}^\eta)^\top \mathbf{B}^\eta \right] \boldsymbol{\eta}^e \ {\rm d}V
\end{aligned}
\end{equation}
For a 2-d model, the matrices appearing in the residuals are defined as follows. The matrix $\mathbf{G}^u$ contains the gradients of the displacement shape functions. For an element with $N_{\text{node}}$ nodes and the nodal displacement vector $\mathbf{u}^e = [u_1^1, u_2^1, ..., u_1^{N_\text{node}}, u_2^{N_\text{node}}]^\top$, $\mathbf{G}^u$ is written as
\begin{equation}
    \mathbf{G}^u = \left[
    \begin{array}{ccccc}
         \partial N^1/\partial X_1  & 0 & ... & \partial N^{N_\text{node}}/\partial X_1  & 0 \\
         0 &  \partial N^1/\partial X_1 & ... & 0 &  \partial N^{N_\text{node}}/\partial X_1 \\
         \partial N^1/\partial X_2  & 0 & ... & \partial N^{N_\text{node}}/\partial X_2  & 0 \\
         0 &  \partial N^1/\partial X_2 & ... & 0 &  \partial N^{N_\text{node}}/\partial X_2 \\
    \end{array}
    \right]
\end{equation}
The first Piola-Kirchhoff stress in Voigt notation has the form of $\{\mathbf{P}\} = [P_{11}, P_{21}, P_{12}, P_{22}]^\top$. The shape function matrices for the displacement and scalar fields take the standard form
\begin{equation}
    \begin{gathered}
        \mathbf{N}^u = \left[
        \begin{array}{ccccc}
             N^1 & 0 & ... & N^{N_\text{node}} & 0 \\
             0 & N^1 & ... & 0 & N^{N_\text{node}} \\
        \end{array}
        \right]\\
        \mathbf{N}^d = \mathbf{N}^\eta = [N^1, N^2, ..., N^{N_\text{node}}]
    \end{gathered}
\end{equation}
The matrices $\mathbf{B}^d$ and $\mathbf{B}^\eta$ consist of the gradients of the scalar shape functions and are given by
\begin{equation}
    \mathbf{B}^d = \mathbf{B}^\eta = \left[
        \begin{array}{ccc}
             \partial N^1/\partial X_1 & ... & \partial N^{N_\text{node}}/\partial X_1 \\
             \partial N^1/\partial X_2 & ... & \partial N^{N_\text{node}}/\partial X_2 \\
        \end{array}
        \right]
\end{equation}

To solve the equations, consistent tangent matrices are obtained by differentiating the residual vectors with respect to the corresponding nodal variables. For the displacement, damage, and interface fields, these matrices are defined as
\begin{equation}
\begin{aligned}
    \mathbf{K}_{uu} 
    &= -\frac{\partial \mathbf{R}_u}{\partial \mathbf{u}^e} 
    = \int_{\Omega_e} \omega(d) (\mathbf{G}^u)^\top 
    \mathbf{A} \mathbf{G}^u \ {\rm d}V \\
    \mathbf{K}_{dd} 
    &= -\frac{\partial \mathbf{R}_d}{\partial \mathbf{d}^e} \\
    &= \int_{\Omega_e} \Bigg[
    -\frac{2G_c(\eta)}{c_0 l_c} (\mathbf{N}^d)^\top \mathbf{N}^d
    + \frac{2G_c(\eta)l_c}{c_0} (\mathbf{B}^d)^\top \mathbf{B}^d 
    + \omega''(d) H^+ (\mathbf{N}^d)^\top \mathbf{N}^d
    \Bigg] \ {\rm d}V \\
    \mathbf{K}_{\eta} 
    &= -\frac{\partial \mathbf{R}_\eta}{\partial \boldsymbol{\eta}^e}
    = \int_{\Omega_e} \left[
    (\mathbf{N}^\eta)^\top \mathbf{N}^\eta 
    + l_i^2 (\mathbf{B}^\eta)^\top \mathbf{B}^\eta
    \right] \ {\rm d}V
\end{aligned}
\end{equation}
where the matrix $\mathbf{A}$ is the Voigt notation of fourth-order tensor $\mathbb{A}$, which relates to the increments of the first Piola-Kirchhoff stress over the increments of the displacement gradient, and has the form of
\begin{equation}
    \mathbb{A}_{ijkl} = \mu\delta_{ik}\delta_{jl} + (\mu - \lambda\log J) F^{-1}_{jk}F^{-1}_{li} + \lambda F^{-1}_{ji} F^{-1}_{lk}
\end{equation}
In the present work, the off-diagonal coupling blocks $\mathbf{K}_{ud}$ and $\mathbf{K}_{du}$ are omitted because a staggered scheme is used to solve for $d$ and $\mathbf{u}$ sequentially. This staggered form improves the robustness of the nonlinear solution procedure. The residual equation for the interface field $\eta$ is independent of $\mathbf{u}$ and $d$. Therefore, once the composite configuration is specified, the interface field is obtained by solving its governing equation with $\eta=1$ prescribed at the interface, and is then kept fixed during the subsequent staggered solution for $d$ and $\mathbf{u}$.

\subsection{ABAQUS UEL Implementation}

The proposed formulation is implemented in the commercial finite element software Abaqus through a user-defined element (UEL) subroutine, following a strategy similar to that of Moln\'ar et al.\cite{Molnar2017}. To solve the interface field, crack phase field, and displacement field, three UEL element types are defined and arranged in a layered manner. At each location, three overlapping elements share the same nodes but carry different sets of degree of freedom (DOF). These layers correspond to the in-plane displacements (DOF 1 and 2), the crack phase field (DOF 3), and the interface field (DOF 4), respectively. In addition, a standard Abaqus element with a UMAT material subroutine is introduced as a fourth, purely post-processing layer to transfer internal variables between layers, such as the values of $d$ and $H^+$, but does not contribute stiffness to any DOF. Based on the residual and tangent matrices derived in the previous subsection, the UEL variables \texttt{RHS} and \texttt{AMATRX} are assembled, and the built-in Abaqus solver is then used to solve the fields.

The overall simulation procedure is organized into two analysis steps, both carried out using Abaqus/Standard. In the first step, the interface field is solved; in the second step, the quasi-static displacement field and crack phase field are solved using Newton-Raphson iterations. After the first step, the resulting $\eta$ is transferred to the displacement and crack phase-field layers. Its value at each integration point is used to evaluate the spatially varying parameters $G_c(\eta)$, $\sigma^*(\eta)$, from which the corresponding $a_1(\eta)$ is determined. Within each time step, the staggered scheme is implemented by alternately solving the equation for $\mathbf{u}$ with $d$ held fixed and the equation for $d$ with $\mathbf{u}$ held fixed. After convergence of the current load increment, the converged $\mathbf{u}$ and $d$ are used as the initial state for the next increment.

\section{Model Verification}

\subsection{One-dimensional Model}

As an initial verification, a one-dimensional bar is considered to examine the consistency of the proposed phase-field formulation between the small- and finite-deformation regimes. In the small-deformation limit, the Neo-Hookean formulation is compared with the corresponding linear-elastic model to verify that the finite-deformation implementation recovers the expected linear response. Additional cases involving larger deformations during fracture are then considered to examine how hyperelastic nonlinearity affects the softening response beyond the small-deformation regime. An asymmetric case is also introduced to show how different bulk fracture properties on the two sides of the interface affect the crack phase-field profile.

\begin{figure}[htpb]
    \centering
    \includegraphics[width=1\linewidth]{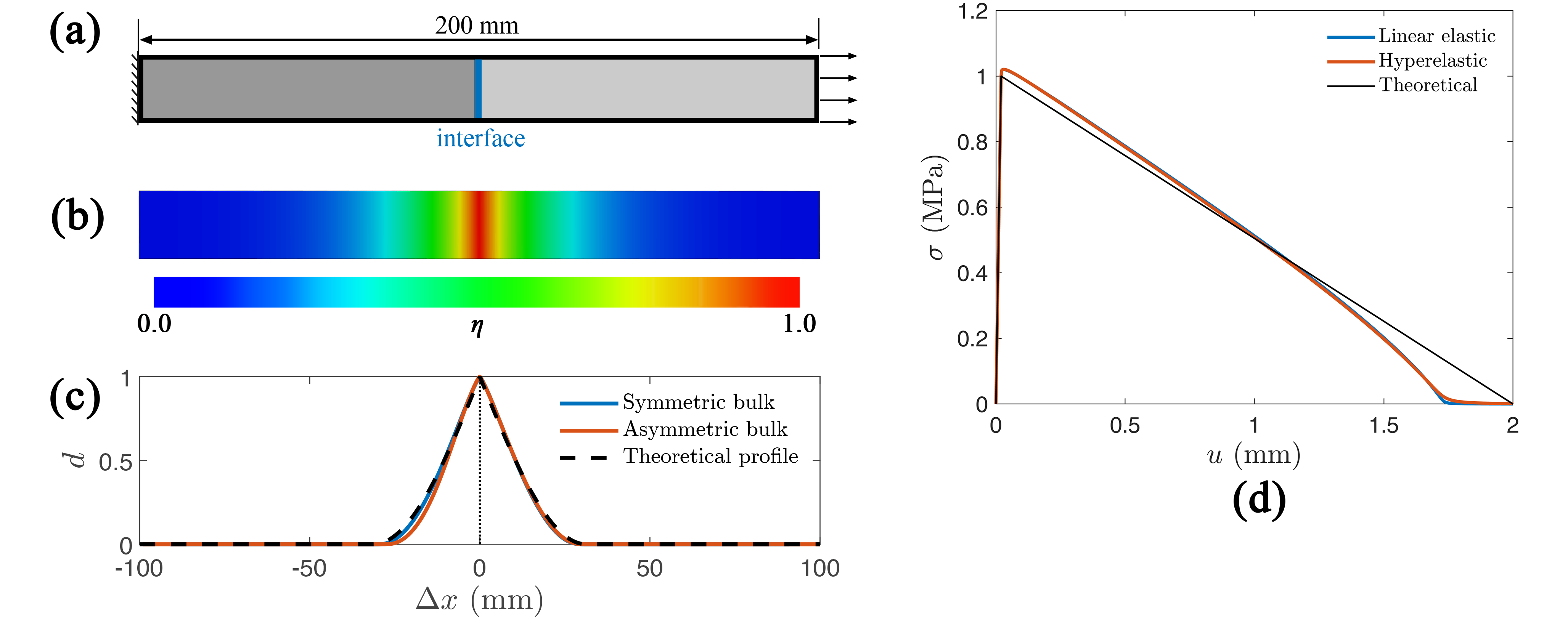}
    \caption{(a) One-dimensional bar with an interface at the middle, subjected to uniaxial tension. (b) Interface field $\eta$. (c) Crack phase-field $d$ at the fully developed fracture state for the symmetric and asymmetric cases. (d) Stress-displacement responses of the linear elastic and Neo-Hookean models, compared with the theoretical $u-\sigma$ response.}
    \label{fig:1d_bar}
\end{figure}

The 1-d bar has a total length of $L = 200 \ \text{mm}$ and contains a single interface at its center, as shown in Fig.~\ref{fig:1d_bar}(a). A monotonic uniaxial displacement is prescribed at one end of the bar, while the other end is kept fixed. The bulk elastic constants are $E = 10^4 \ \text{MPa}$ and $\nu = 0$. A symmetric case with identical bulk fracture properties on both sides of the interface, and an asymmetric case with higher fracture resistance on the left side are considered.  In the symmetric case, the critical tensile strength and critical energy release rate on both sides are set to $\sigma^*_m = 1.25 \ \text{MPa}$ and $G_m = 1.5 \ \text{N/mm}$, respectively. In the asymmetric case, the properties on the right side remain unchanged, whereas those on the left side are doubled to $\sigma^*_m = 2.5 \ \text{MPa}$ and $G_m = 3.0 \ \text{N/mm}$. The interface parameters are $\sigma^*_i = 1.0 \ \text{MPa}$ and $G_i = 1.0 \ \text{N/mm}$ for both cases. The characteristic length scales of the crack and interface fields are set to $l_c = l_i = 20\ \text{mm}$. The bar is discretized with a uniform mesh size of $h = 0.1 \ \text{mm}$, and the end displacement is increased in increments of $\Delta u = 4\times10^{-6} \ \text{mm}$.

For the interface field, Fig.~\ref{fig:1d_bar}(b) shows the distribution of $\eta$, which is equal to 1 at the interface and decays to 0 on both sides. The crack phase-field profiles of the Neo-Hookean system at the fully developed fracture state are compared in Fig.~\ref{fig:1d_bar}(c), where $\Delta x$ denotes the coordinate relative to the crack center. The dashed curve represents the theoretical 1-d profile derived for a spatially uniform fracture energy. Even for the symmetric case, the simulated profile is slightly narrower than the theoretical profile because the fracture properties vary continuously through $\eta$, from the lower $G_i$ near the interface toward the higher $G_m$. When the fracture resistance on the left side is increased, $d$ profile becomes asymmetric, with a faster decay toward the side with the higher bulk strength and fracture energy. This behavior is consistent with previous studies showing that spatial variations in $G_c$ modify $d$ profile and consequently introduce a deviation between the prescribed $G_i$ and the actual fracture dissipation, which depends on the ratio of length scale $l_i/l_c$ \cite{Yoshioka2021,Zhou2022,Bian2025,Chen2025}. This deviation mainly affects the quantitative calibration of $\tilde{G}_i$ rather than the applicability of the diffuse-interface formulation, and can be further reduced through energy-consistent corrections\cite{Zhou2022,Chen2025}.

Fig.~\ref{fig:1d_bar}(d) compares the nominal stress-displacement responses of the linear-elastic and Neo-Hookean models for the symmetric reference case, with the black curve representing the theoretical $u-\sigma$ response, including both the elastic deformation of the bulk and the bilinear interfacial softening. The two simulation results produce nearly identical curves, and both show approximately 2\% higher peak stress and a slightly smaller failure displacement than the theoretical prediction. This small deviation is associated with the diffuse representation of the interface and has been discussed in previous work\cite{Zhen2024}, who attributed it to the higher critical energy release rate over most of the crack-localization band compared with that at the interface. Overall, the overlap between the linear-elastic and hyperelastic responses indicates that the Neo-Hookean system recovers the linear-elastic response in the small-deformation limit.

As the deformation increases, the Neo-Hookean response gradually deviates from the linear-elastic relation. We therefore perform an additional verification at larger strain levels. Starting from the baseline setup, a scaling factor $n$ is introduced to generate a family of systems with larger fracture strains while keeping the modulus unchanged. The matrix and interface fracture parameters are scaled by the same factor. In particular, the tensile strengths are scaled as $\sigma^*_m(n) = n \sigma^*_{m0}$ and $\sigma^*_i(n) = n \sigma^*_{i0}$, and the critical energy release rates are scaled as $G_m(n) = n^2 G_{m0}$ and $G_i(n) = n^2 G_{i0}$. Here, the subscript 0 denotes the reference system with $n=1$, which uses the baseline parameters given above. Under the ideal bilinear softening law derived for a linear-elastic bulk, this scaling implies that both the peak stress and the theoretical failure displacement $u^{\rm theo}_f$ increase by a factor of $n$, with $u^{\rm theo}_f(n) = n\,u^{\rm theo}_{f0}$ and $u^{\rm theo}_{f0} = 2\ \text{mm}$. Therefore, the normalized curves in terms of $\sigma/\sigma^*_i$ and $u/u^{\rm theo}_f$ are expected to overlap.

\begin{figure}[htpb]
    \centering
    \includegraphics[width=1.0\linewidth]{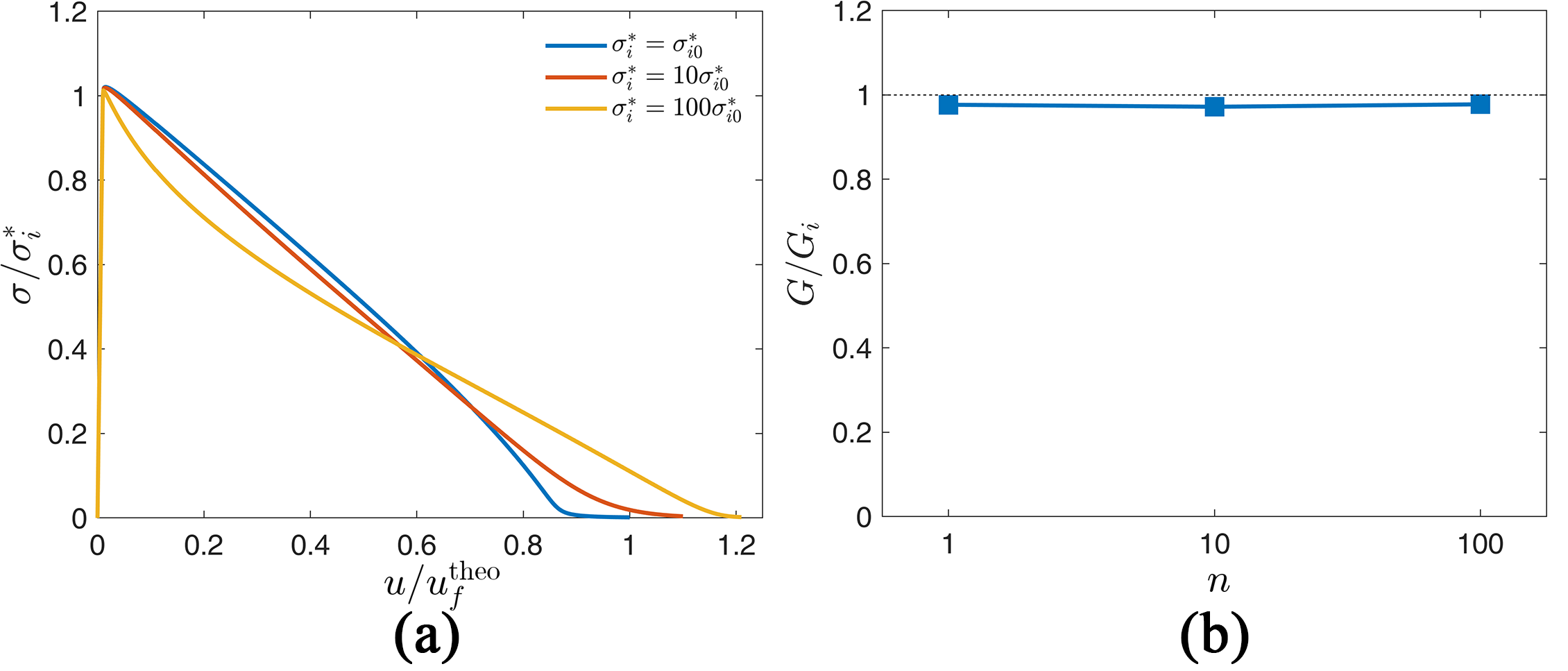}
    \caption{(a) Normalized stress-displacement responses of the 1-d bar for different scaling factors. (b) Ratio of the dissipated energy $G$ to the target value $G_i$ for the same set of scaling factors.}
    \label{fig:1d_hyperelastic}
\end{figure}

Here, the scaling parameter is set to $n = 10$ and $n = 100$, and the simulated normalized stress-displacement responses are shown in Fig.~\ref{fig:1d_hyperelastic}(a). As $n$ increases, the response deviates from the bilinear reference model. Compared with the baseline case, the normalized stress-displacement curves become convex and exhibit a longer tail before complete failure. This deviation is expected because the bilinear reference is based on linear elasticity, whereas a hyperelastic bulk exhibits a nonlinear response at large strains and therefore does not preserve the ideal bilinear shape in the normalized plot. To assess whether the total dissipation still matches the prescribed fracture energy, the dissipated energy is computed by integrating the force-displacement response. Fig.~\ref{fig:1d_hyperelastic}(b) shows the ratio of the simulated dissipation $G$ to the target value $G_i$. The normalized dissipation $G/G_i$ remains close to 1, with a slight underestimation, and is nearly unchanged for different values of $n$. Therefore, although the large-strain hyperelastic response modifies the softening shape, it does not introduce an additional deviation from the prescribed fracture energy.

Overall, the 1-d verification confirms that the proposed phase-field framework can consistently describe interfacial fracture in a hyperelastic bulk with a smeared interface field, supporting its use in the subsequent simulations of weak-interface effects in particle-filled polymers.

\subsection{Single Fiber System under Transverse Tension}

\begin{figure}[htpb]
    \centering
    \includegraphics[width=0.4\linewidth]{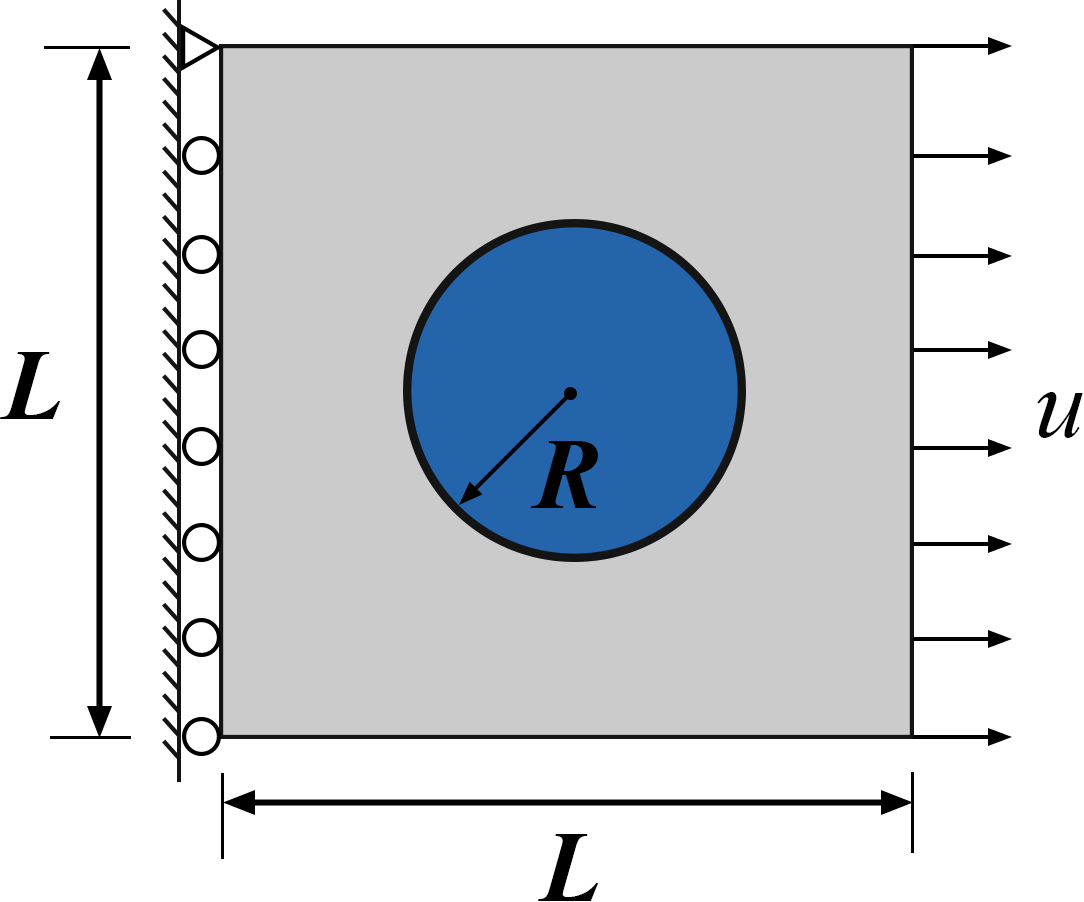}
    \caption{Single-fiber model. A circular fiber of radius $R$ is centered in a square matrix of edge length $L$. The left edge is constrained in the loading direction, and a horizontal displacement $u$ is applied on the right edge.}
    \label{fig:SingleInclusionModel}
\end{figure}

We next simulate a benchmark model consisting of a square matrix with a single fiber embedded at its center under transverse tension. The geometry and boundary conditions are shown in Fig.~\ref{fig:SingleInclusionModel}. The matrix has an edge length of $L = 1 \ {\rm mm}$, and the fiber radius is $R = 0.25 \ {\rm mm}$. The matrix is modeled using Neo-Hookean hyperelasticity, while the fiber is assumed to be linear elastic. Following previous studies\cite{Zhen2024}, the Young's modulus and Poisson's ratio are set to $E_m = 4\times10^3 \ {\rm MPa}$ and $\nu_m = 0.4$ for the matrix, and to $E_{\rm fiber} = 4\times10^4 \ {\rm MPa}$ and $\nu_{\rm fiber} = 0.33$ for the fiber. For fracture properties, the critical strength and critical energy release rate are set to $\sigma^*_m = 30 \ {\rm MPa}$ and $G_m = 0.25 \ {\rm N/mm}$, respectively, and the corresponding interface parameters are $\sigma^*_i = 10 \ {\rm MPa}$ and $G_i = 0.05 \ {\rm N/mm}$. The characteristic length scales are set to $l_i = 0.015 \ {\rm mm}$ for the interface field and $l_c = 0.010 \ {\rm mm}$ for the crack phase field. A refined mesh is used around the interface and in the surrounding matrix region where crack kinking may occur, with an element size of $h = 2\times 10^{-3} \ {\rm mm}$. Displacement loading is applied to the boundary, and the imposed displacement is increased by $\Delta u = 2\times10^{-6} \ {\rm mm}$ per increment.

\begin{figure}[htpb]
    \centering
    \includegraphics[width=1.0\linewidth]{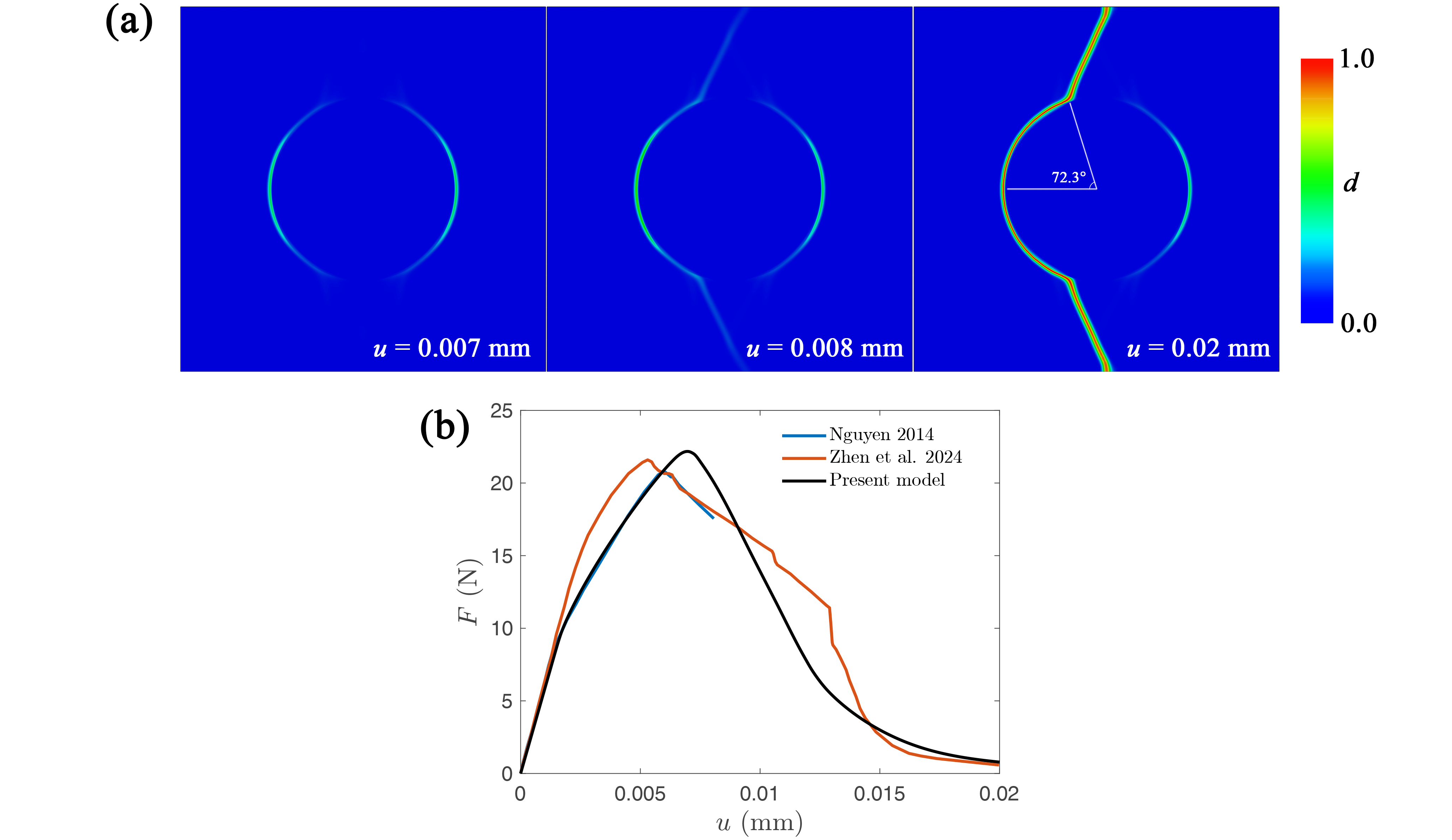}
    \caption{(a) Crack phase-field $d$ development in the single-fiber system at different applied displacements, showing interfacial cracking followed by kinking into the matrix. (b) Force-displacement response of the present model compared with the literature results\cite{Nguyen2014,Zhen2024}.}
    \label{fig:SingleInclusion}
\end{figure}

The simulation results for the single-fiber system are shown in Fig.~\ref{fig:SingleInclusion}. The phase-field snapshots in Fig.~\ref{fig:SingleInclusion}(a) show the characteristic failure process associated with a weak interface. Damage initiates at the interface, propagates along it over a limited distance, and subsequently kinks into the matrix, leading to final failure. Since the deformation in this benchmark remains relatively small, the comparison is used as a limiting case verification of whether the present finite-deformation formulation recovers the interfacial fracture behavior. The kinking angle is commonly used as a quantitative measure of the interface-to-matrix crack transition. The kinking angle obtained in the present simulation is $72.3^\circ$, which is slightly larger than the values reported in earlier studies\cite{Nguyen2014,Zhen2024}. This relatively small difference may be attributed to the different bulk constitutive descriptions, because the hyperelastic response can slightly change the local stress state near the interface compared with the linear-elastic response in benchmark references.

In addition to the crack path, Fig.~\ref{fig:SingleInclusion}(b) compares the force-displacement response with previously reported results\cite{Nguyen2014,Zhen2024}. The overall curve shape and peak stress level agree with the reference results, with only minor deviations. Overall, the agreement in the failure mode, kinking behavior, and global response provides a consistency check of the present formulation in the small-deformation limit.

\subsection{RVE Model}

\subsubsection{Construction of RVE Model}

This study focuses on the mechanical response and fracture process of particle-filled composites. To represent the behavior of the heterogeneous material, a 2-d square RVE with size $L \times L$ is adopted under plane-strain conditions. The particles are modeled as circular particles with a uniform radius $R$ and are placed randomly inside the RVE to reach the target filler volume fraction, which is represented by the particle area fraction in the 2-d model. In the simulations, the particle radius is set to $R = 0.17\ {\rm mm}$. To avoid extremely thin gaps between neighboring particles and the associated meshing difficulties, the minimum edge-to-edge distance between any two particles is set to $\ge 0.05 R$.

\begin{figure}[htpb]
    \centering
    \includegraphics[width=0.8\linewidth]{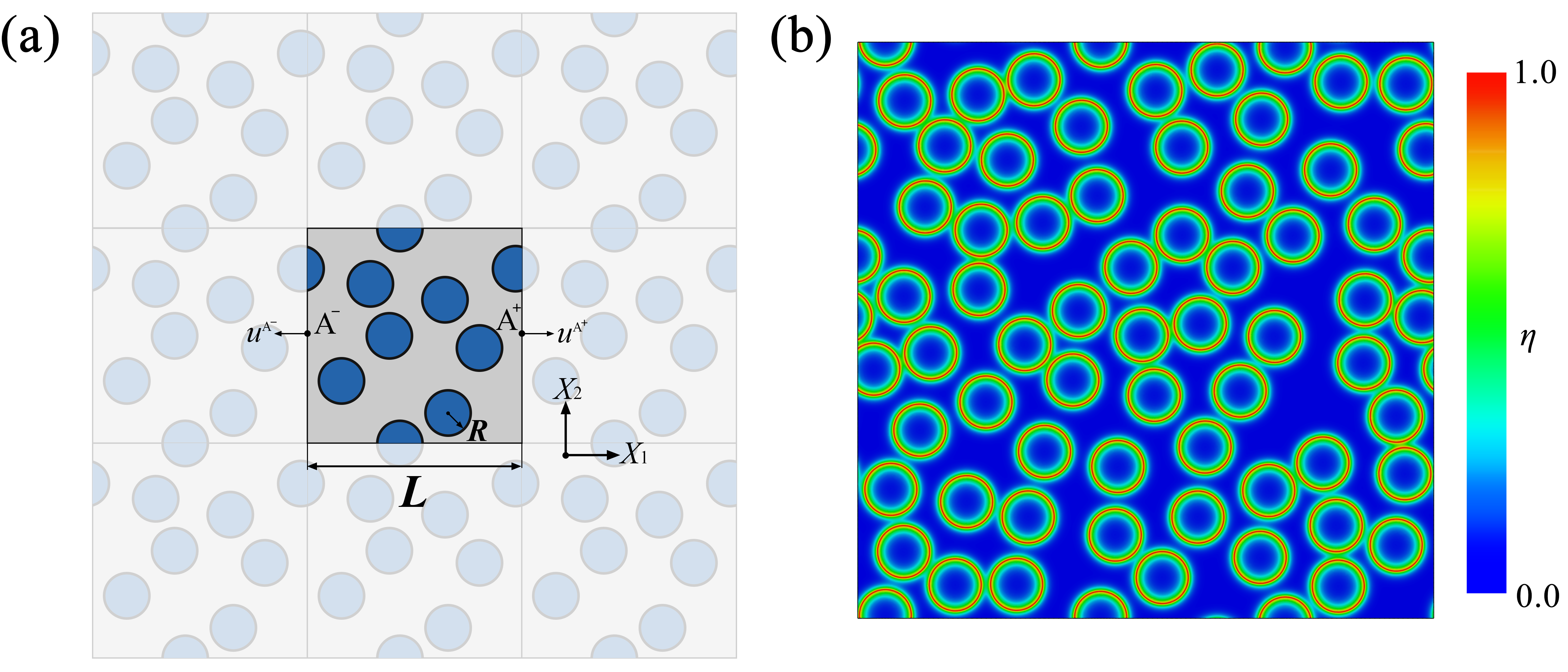}
    \caption{(a) Schematic of the representative volume element (RVE) with periodic boundary condition (PBC). The RVE has size $L \times L$ and contains circular particles of radius $R$. Under PBC, the displacement relation between two opposite nodes $u_i^{A^\pm}$ is determined by the average strain. (b) Spatial distribution of the interface field $\eta$ in the RVE.}
    \label{fig:PBCandEta}
\end{figure}

For the boundary conditions, periodic boundary conditions (PBC) are applied to relate the fields on opposite boundaries of the RVE. Consider a pair of corresponding nodes on two opposite boundaries of the RVE, denoted by $\text{A}^+$ and $\text{A}^-$, with initial positions $\mathbf{X}^{A^+}$ and $\mathbf{X}^{A^-}$. Under PBC, the interface field and crack phase field satisfy
\begin{equation}
    \eta^{A^+}=\eta^{A^-}, \qquad d^{A^+}=d^{A^-},
\end{equation}
whereas the displacement relation between the two nodes is determined by the average strain $\bar{\varepsilon}_{ij}$ as
\begin{equation}
    u_i^{A^+} - u_i^{A^-}
    = \bar{\varepsilon}_{ij} (X_j^{A^+} - X_j^{A^-})
    = \bar{\varepsilon}_{ij} \Delta X_j^0,
\end{equation}
where $u_i^{A^\pm}$ denotes the displacement of node $\text{A}^\pm$ in the $i$ direction, and $\Delta X_j^0 = X_j^{A^+} - X_j^{A^-}$ is the initial distance between the two nodes in the $j$ direction. The overall configuration is illustrated in Fig.~\ref{fig:PBCandEta}(a), and the representative spatial distribution of the interface field $\eta$ is shown in Fig.~\ref{fig:PBCandEta}(b).

For uniaxial stretching along the $X_1$ direction, the macroscopic axial strain $\bar{\varepsilon}_{11}$ is controlled by the displacement constraints. The macroscopic transverse nominal stress is set to zero, and the transverse strain $\bar{\varepsilon}_{22}$ is left unconstrained and determined from the equilibrium solution. In the simulation, loading is applied in a displacement-controlled manner. Before damage initiation, the displacement increment on the RVE boundary is set to $\Delta u = 3\times 10^{-2}\ {\rm mm}$ per load step, and once fracture starts to develop, the increment is decreased to $\Delta u = 1\times 10^{-3}\ {\rm mm}$ to accurately capture crack evolution. The mesh size in the polymer matrix is $h = 8\times 10^{-3}\ {\rm mm}$.

The macroscopic nominal axial stress is calculated from the total axial reaction force on the loaded boundary of the RVE. In the present 2-d setting with unit thickness, the nominal axial stress is defined as $\bar{\sigma}^{\rm nom}_{11} = F_1/L$, where $F_1$ is the total reaction force in the loading direction $X_1$, obtained by summing the nodal reaction forces on the loaded boundary. In the following analysis, the macroscopic stress-strain curves are presented in terms of the nominal axial stress $\bar{\sigma}^{\rm nom}_{11}$ (denoted by $\sigma$) and the applied axial engineering strain $\bar{\varepsilon}_{11}$ (denoted by $\varepsilon$). Accordingly, the resulting RVE curve is interpreted as the effective material response under macroscopically homogeneous uniaxial deformation and can therefore be compared with the nominal stress-strain curve obtained from tensile tests.

\subsubsection{RVE Representativeness and Sensitivity Analyses}

To assess whether the RVE model provides sufficiently representative results for the subsequent analyses, the effects of particle configuration, RVE size, characteristic length scale and mesh size are investigated.

In the verification simulations, the filler volume fraction is fixed at $f = 0.42$. The polymer matrix is modeled as a hyperelastic material with Lam\'e parameters $\mu = 1.16\ {\rm MPa}$ and $\lambda = 0.775\ {\rm MPa}$, calibrated from the tensile response of the unfilled PU rubber \cite{Schwarzl1967}. The particles are modeled as a linear-elastic material with Young's modulus $E = 2000\ {\rm MPa}$ and Poisson's ratio $\nu = 0.15$.
Since the present study focuses on interfacial debonding and matrix fracture, the particle tensile strength and critical energy release rate are set to $\sigma^*_p = 5.0\ {\rm MPa}$ and $G_p = 1.0\ {\rm N/mm}$, to suppress particle fracture, similar to previous phase-field studies\cite{Zhen2024,NguyenThanh2022}. The critical energy release rates of the polymer matrix and interface are $G_m = 0.56\ {\rm N/mm}$ and $G_i = 0.23\ {\rm N/mm}$, respectively. The tensile strengths are set to $\sigma^*_m = 1.25\ {\rm MPa}$ for the polymer matrix and $\sigma^*_i = 0.42\ {\rm MPa}$ for the interface. These parameters define an elastomer composite system in which the polymer matrix has low stiffness but can sustain large deformations before damage initiation and fracture, whereas the particles are much stiffer and the particle-matrix interfaces are relatively weak. The parameter values are selected with reference to the experimentally calibrated composite system presented in the following section.

\begin{figure}[htpb]
    \centering
    \includegraphics[width=1.0\linewidth]{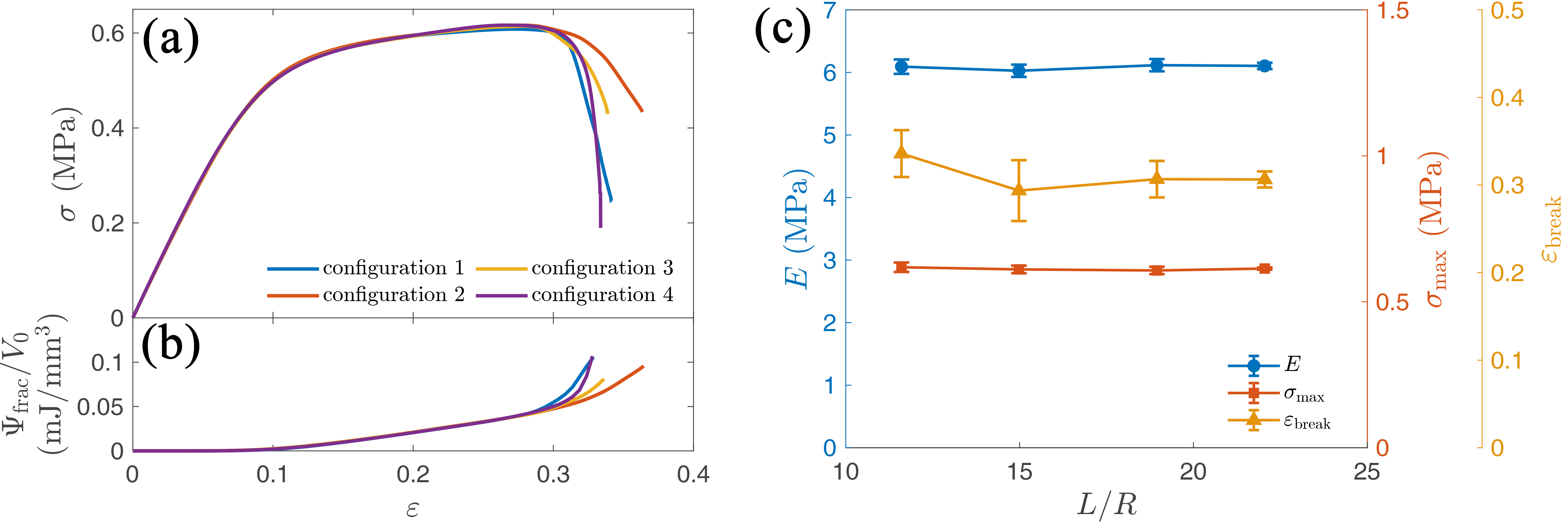}
    \caption{ (a) Macroscopic stress-strain responses of four random particle configurations at $L = 22.0 R$. (b) Accumulated fracture energy per unit volume $\Psi_{\rm frac}/V_0$ of four random particle configurations at $L = 22.0 R$. (c) Mean value and standard deviation of initial modulus $E$, maximum stress $\sigma_{\rm max}$, and the strain at break $\varepsilon_{\rm break}$ for different RVE sizes simulated from 4 different configurations.} 
    \label{fig:RVEsensiti}
\end{figure}

To assess the effect of random particle arrangement, we compare the mechanical responses of RVE with different particle configurations. The length scales are fixed at $l_i = l_c = 0.02\ \text{mm}$, and the RVE side length is $L = 22.0 R$. Four randomly packed configurations are generated at the same filler volume fraction and RVE size, and their tensile deformation and fracture processes are simulated. The corresponding stress-strain curves are shown in Fig.~\ref{fig:RVEsensiti}(a). Up to the abrupt stress drop, the response can be interpreted as the effective tensile behavior of the composite, whereas the abrupt drop marks fracture localization and corresponds to macroscopic fracture in a tensile specimen. The response up to this point captures the evolution from deformation and interfacial debonding to damage coalescence and dominant fracture-band formation, whereas the residual post-drop response is of secondary importance for characterizing the tensile behavior and failure of the composite. To further compare the fracture evolution, the accumulated fracture energy is calculated as $\Psi_{\rm frac}=\int_{\Omega} G_c(\eta)\gamma(d,\nabla d)\,{\rm d}V$ and normalized by the initial RVE volume $V_0$, as shown in Fig.~\ref{fig:RVEsensiti}(b). The four configurations exhibit nearly identical dissipation energy profiles before fracture localization, and the rapid increase in fracture energy occurs at similar strain, coinciding with the abrupt stress drop. Together with the similar stress-strain responses, these results indicate that random particle arrangement has a limited influence on the response of the selected RVE.

The influence of RVE size on the predicted response is then investigated. Four RVE side lengths $11.6R$, $15.0R$, $18.9R$, and $22.0R$ are considered, and four random particle configurations are generated for each size. Since the RVE analysis aims to characterize the effective response up to the abrupt stress drop, representativeness is evaluated using the initial modulus $E$, maximum stress $\sigma_{\rm max}$, and strain at break $\varepsilon_{\rm break}$. Here, $\varepsilon_{\rm break}$ is defined as the nominal strain at the onset of the abrupt stress drop and is used as a numerical measure associated with the experimentally observed elongation at break. For different RVE sizes, these three characteristic parameters are extracted and their standard deviations are analyzed, as shown in Fig.~\ref{fig:RVEsensiti}(c).

The results show that, as the RVE size increases, $E$ and $\sigma_{\rm max}$ are nearly unchanged, indicating insensitivity to RVE size and their standard deviations also remain small. Before the abrupt stress drop, the macroscopic response is dominated by the nonlinear elastic behavior of the polymer matrix, and damage initiates and spreads diffusely throughout the domain. Because no clear localization band has formed at this stage, increasing the RVE size mainly repeats the similar deformation and damage pattern. Therefore the mechanical response in this regime is essentially independent of RVE size. In contrast, $\varepsilon_{\rm break}$ is more sensitive to RVE size, and its deviation is relatively large when $L$ is small. This size dependence arises because after the abrupt stress drop associated with localized fracture-band formation, the width of the crack band is governed by the characteristic crack-band length scale and does not scale with the RVE size. For smaller RVEs, the post-peak softening is less steep, which may lead to a slight increase in $\varepsilon_{\rm break}$. However, when the RVE is sufficiently large ($L \ge 18.9 R$), $\varepsilon_{\rm break}$ approaches a constant value. Accordingly, $L = 22.0R$ is adopted in the following study, for which the response is less sensitive to RVE size and particle configuration and the deviation is smaller.

\begin{figure}[htpb]
    \centering
    \includegraphics[width=1.0\linewidth]{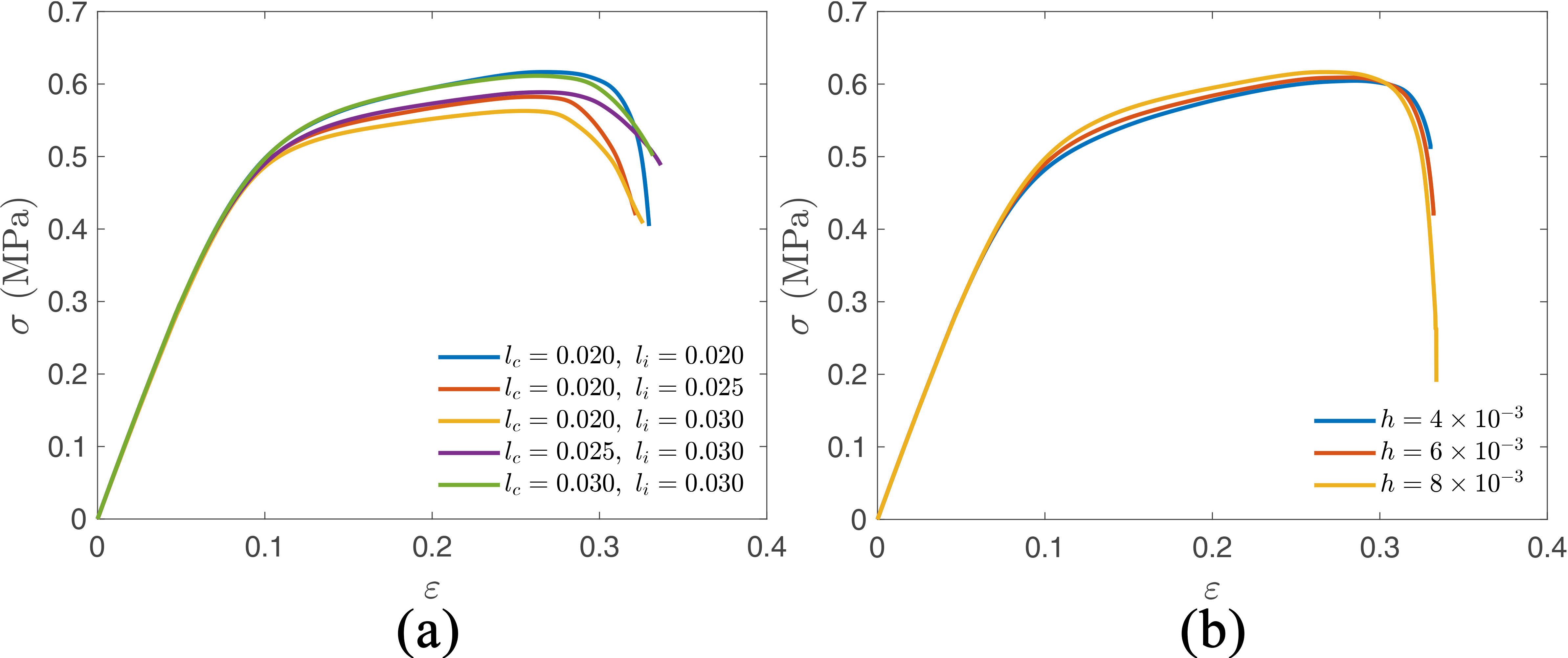}
    \caption{Sensitivity of the macroscopic stress-strain response to the phase-field length scales and mesh size. (a) Effects of the crack length scale $l_c$ and interface length scale $l_i$. (b) Effects of mesh size $h$.}
    \label{fig:sensitiLcLi}
\end{figure}

Then, the effects of the characteristic length scales $l_c$ and $l_i$ are assessed, and the corresponding stress-strain curves are shown in Fig.~\ref{fig:sensitiLcLi}(a). With the crack length scale fixed at $l_c = 0.02\ {\rm mm}$, increasing $l_i$ from $0.02\ {\rm mm}$ to $0.03\ {\rm mm}$ produces slight decreases in both peak stress and corresponding strain. With the interface length scale fixed at $l_i = 0.03\ {\rm mm}$, increasing $l_c$ from $0.02\ {\rm mm}$ to $0.03\ {\rm mm}$ leads to a slight increase in peak stress and fracture strain. These variations indicate that the present formulation has a weak dependence on the regularization lengths. In the present model, the effective interfacial toughness $\tilde{G}_i$ is determined from a 1-d calibration so that the total fracture energy dissipated across the diffuse interface equals prescribed $G_i$. Such corrections can reduce the dependence on the regularization lengths, but some residual dependence may remain in higher-dimensional or heterogeneous problems because the spatial variation of $G_c$ may cause the actual crack phase-field distribution to deviate from the ideal 1-d profile \cite{HansenDrr2019,Zhou2022,Chen2025}. More rigorous energy-consistent corrections considering the deviation of phase-field profile have been proposed to further reduce this dependence \cite{Yoshioka2021,Zhou2022,Chen2025,Bian2025}. Although some residual dependence remains, it is substantially reduced compared with the stronger length-scale sensitivity of the apparent tensile strength in the standard AT2 formulation \cite{NajafiKoopas2025}. Within the systems considered here, the remaining shifts are relatively small and have little effect on the overall response characteristics or the qualitative conclusions of the subsequent analyses.

A mesh-convergence study is also performed, and the corresponding stress-strain curves are shown in Fig.~\ref{fig:sensitiLcLi}(b). As the mesh size increases from $h=4\times10^{-3}\ {\rm mm}$ to $8\times10^{-3}\ {\rm mm}$, the curves remain nearly unaffected, with only a slight increase in the maximum stress and a small decrease in the failure strain, indicating that the macroscopic response is only weakly affected by further mesh refinement and that the adopted mesh size provides sufficient numerical convergence.

Based on the above sensitivity analyses of RVE size, particle configuration, and characteristic length scales, the macroscopic stress-strain response of the composite is only weakly affected by variations in these parameters. This confirms that the chosen RVE setting provides a robust basis for the subsequent studies. In all subsequent simulations, the RVE side length is fixed at $L = 22.0R$, mesh size is fixed at $h=8\times10^{-3}\ {\rm mm}$, and the length scales are set to $l_c = l_i = 0.02\ {\rm mm}$.

\section{Experimental Comparison and Effects of Filler Volume Fraction}

\subsection{Comparison with Experimental Stress-Strain Curves at Different Filler Volume Fractions}

\begin{figure}[htpb]
    \centering
    \includegraphics[width=0.6\textwidth]{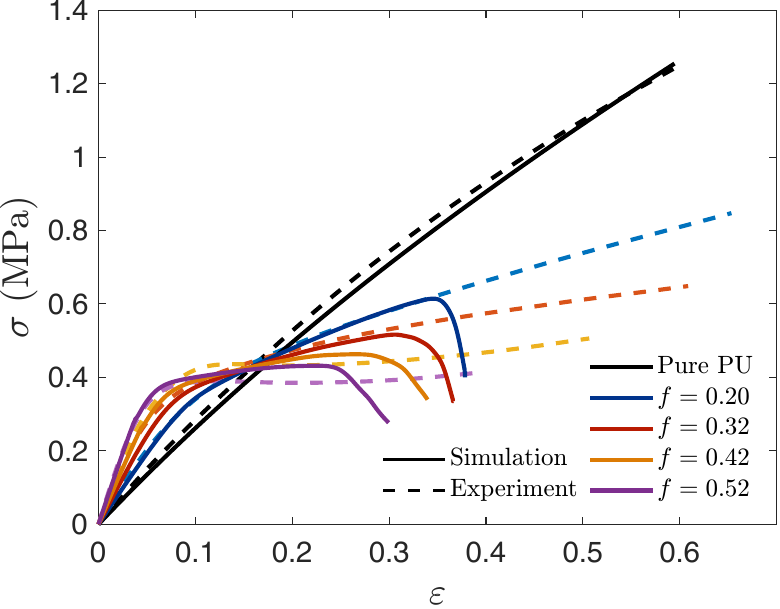}
    \caption{Experimental and simulated stress-strain curves of polyurethane rubber composites with different sodium chloride filler volume fractions $f$. The solid lines represent the simulation results, whereas the dashed lines represent the experimental data.}
    \label{fig:solid_content_compare}
\end{figure}

To investigate how weak interface affects the macroscopic mechanical response and damage evolution of particle-filled polymer composites, a common set of model parameters is first calibrated by comparing the predicted responses with experimental stress-strain curves for polyurethane (PU) rubber filled with increasing volume fractions of coarse sodium chloride particles ($210$--$300\,\mu\mathrm{m}$)\cite{Schwarzl1967}. The unfilled PU rubber exhibits a hyperelastic response, which is described using the Neo-Hookean model with Lam\'e parameters $\mu = 1.16\ {\rm MPa}$ and $\lambda = 0.775\ {\rm MPa}$. Based on the experimental response of the unfilled PU rubber, the critical strength of the matrix is set to $\sigma^*_m = 1.25\ {\rm MPa}$. The experimental response of the unfilled PU rubber and the corresponding Neo-Hookean fit are shown as the black curves in Fig.~\ref{fig:solid_content_compare}. With the matrix constitutive parameters and critical strength determined, the remaining interfacial and fracture parameters are calibrated by considering the main features of the experimental curves across all filler volume fractions. The interfacial strength is chosen mainly according to the stress level at the onset of softening, while the fracture-energy parameters are adjusted to provide a representative description of the post-softening response. The resulting parameters are $\sigma^*_i = 0.25\ {\rm MPa}$, $G_m = 0.56\ {\rm N/mm}$, and $G_i = 0.23\ {\rm N/mm}$. The same parameter set is used for all filler volume fractions, rather than fitting each experimental curve separately.

As shown in Fig.~\ref{fig:solid_content_compare}, the simulations reproduce the effective initial modulus, the onset of softening, and the post-softening tangent modulus well. As the filler volume fraction increases, the effective initial modulus increases because of the reinforcing effect of the stiff filler particles. In contrast, the softening stress remains nearly unchanged, as it is mainly governed by the debonding strength of the polymer--filler interface. The post-softening tangent modulus decreases with increasing filler volume fraction because a larger interfacial area is involved in the damage and softening process. The simulations also reproduce the experimental trend that both the stress and strain at break decrease with increasing filler volume fraction, although their absolute values are underestimated.

Using a single parameter set for all filler volume fractions, the model captures the main features of the experimental curves and their dependence on filler volume fraction, despite the simplifications introduced by the 2-d representation and the idealized circular particle geometry. The 2-d representation simplifies the 3-d geometry and changes the total particle-matrix interfacial area, which may affect the debonding process and lead to quantitative deviations in the post-softening response, particularly at high filler volume fractions. Particle morphology can also modify the local stress distribution, damage development, and macroscopic fracture response of particle-filled composites, as demonstrated by previous experimental and numerical studies comparing spherical and faceted or polyhedral particles\cite{deFrancqueville2020_2,Weng2019,Yang2022}. Accordingly, representing the faceted NaCl particles by idealized circular geometries may introduce quantitative deviations in the predicted mechanical and fracture responses. Besides, the underestimation in the final failure stage may partly arise from directly using the macroscopic failure strength measured for the unfilled PU rubber as the local matrix damage-initiation strength in the micromechanical model. Because these quantities characterize failure at different physical scales and under different deformation conditions, directly adopting the experimental value as a phase-field parameter may lead to the quantitative deviation in the predicted failure stress and strain. The choice of interfacial parameters may also contribute to the discrepancy, since the interfacial fracture energy $G_i$ affects the damage evolution after damage initiation, including progressive debonding and fracture localization. Overall, the comparison indicates that the present model provides a suitable basis for investigating the mechanical response and fracture process of particle-filled polymer composites.

\subsection{Strength Trends for Weak-Interface and Well-Bonded Systems}

To illustrate the effect of interfacial weakening on the filler-volume-fraction dependence of the composite response, a well-bonded reference system is considered, in which the interfacial strength and fracture energy are set equal to the matrix, namely $\sigma^*_i = \sigma^*_m$ and $G_i = G_m$. This eliminates the contrast in fracture properties between the interface and the matrix, so that the particle boundary no longer acts as a preferential path for interfacial debonding. Consequently, any damage developing near the particle boundary is governed by the matrix fracture properties and is interpreted as matrix cracking rather than interfacial failure. Within the present phase-field framework, this setting therefore represents an ideally bonded particle--matrix interface. All other matrix and particle parameters are kept identical to those used in the preceding experimental comparison.

\begin{figure}[htpb]
    \centering
    \includegraphics[width=1.0\textwidth]{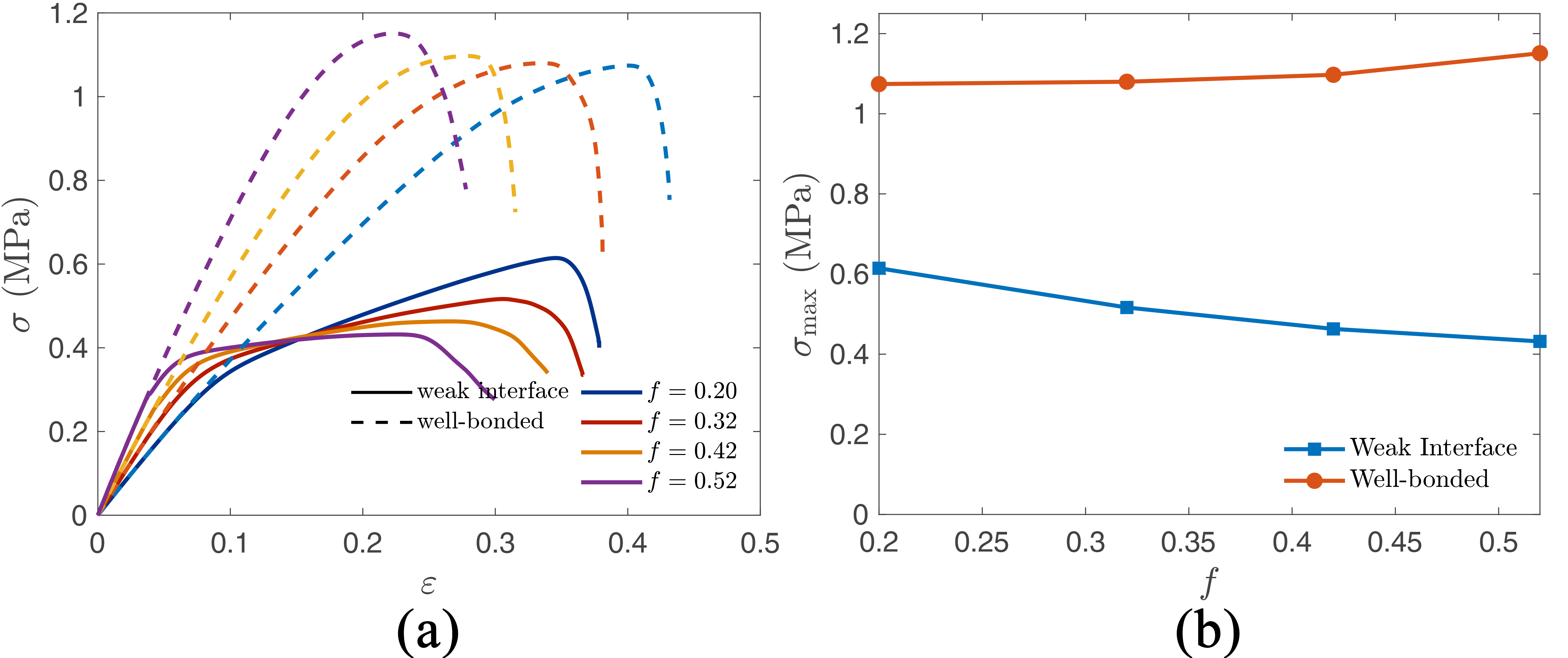}
    \caption{Comparison of the weak-interface and well-bonded systems. (a) Simulated stress-strain curves. The solid lines correspond to the calibrated weak-interface parameters, whereas the dashed lines are obtained by setting the interfacial strength and fracture energy equal to the polymer matrix, representing the well-bonded reference system. (b) Maximum nominal stress $\sigma_{\rm max}$ for the weak-interface and well-bonded systems at different filler volume fractions $f$.}
    \label{fig:solid_content_IntactSurf}
\end{figure}

The simulated stress-strain curves are shown in Fig.~\ref{fig:solid_content_IntactSurf}(a). Removing the weak interface largely changes the overall response. In the weak-interface system, the stress-strain curve exhibits a distinct softening point followed by an interface-controlled regime. However, in the well-bonded system this regime disappears, and the stress continues to increase until final failure. As a result, the maximum stress is much higher than that in the weak-interface system. Besides, removing interfacial weakening also leads to a slight increase in the strain at break.

The two systems exhibit different trends in maximum stress with increasing filler volume fraction. In the weak-interface system, increasing the filler volume fraction increases the total area of weak interfaces and promotes more extensive interfacial debonding, thereby reducing the maximum stress. In the well-bonded system, the maximum stress is less sensitive to filler volume fraction and increases slightly because of the reinforcing effect of the stiff fillers. These contrasting trends are shown in Fig.~\ref{fig:solid_content_IntactSurf}(b). This comparison helps explain the different strength trends reported for particle-filled polymer composites. Previous studies have shown that the strain at break generally decreases with increasing filler volume fraction, whereas the tensile strength may either increase or decrease depending on the filler type\cite{Fu2008} and interfacial treatment\cite{Nunes2000}. These observations suggest that interfacial properties should be considered when evaluating the effect of filler volume fraction, because they can lead to opposite trends in macroscopic strength.

\section{Multi-Stage Fracture Induced by Interfacial Weakening}

\subsection{Influence of Interfacial Strength on Multi-Stage Fracture}

\begin{figure}[htpb]
    \centering
    \includegraphics[width=0.6\linewidth]{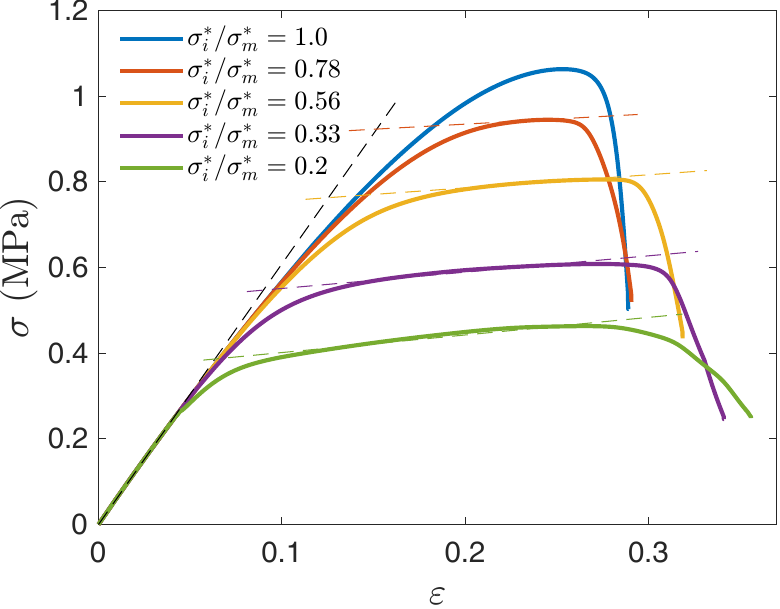}
    \caption{Effect of the interfacial-to-matrix strength ratio $\sigma^*_i/\sigma^*_m$ on the macroscopic stress-strain response of the composite. Dashed lines indicate linear fits to the initial undamaged regime and the intermediate softening regime.}
    \label{fig:diff_sigmai}
\end{figure}

To further systematically study how weak interface affects the mechanical response and fracture process, a system based on the previously calibrated parameters is considered, in which the interfacial strength is varied. The filler volume fraction is fixed at $f = 0.42$. The elastic properties of the polymer matrix and particles, the critical energy release rates of the matrix and interface, and the tensile strength of the polymer matrix are kept the same as those in the previous section, while the interfacial strength $\sigma^*_i$ is gradually reduced from $\sigma^*_i = \sigma^*_m$ to $\sigma^*_i = 0.20\sigma^*_m$. The corresponding stress-strain curves are shown in Fig.~\ref{fig:diff_sigmai}.

The simulated stress-strain curves show that when the critical strengths of the matrix and the interface are equal, i.e. $\sigma^*_m / \sigma^*_i = 1$, the composite response initially follows the nonlinear elastic behavior of the polymer matrix, and after damage initiation, the curve gradually softens to a stress peak, followed by an abrupt stress drop that marks the final fracture of the composite. When the interfacial strength $\sigma^*_i$ is reduced below the matrix strength, an additional stable softening regime gradually appears between the undamaged regime and the final abrupt failure. Within this intermediate regime, the macroscopic stress-strain response is approximately linear, as indicated by the dashed linear fits in Fig.~\ref{fig:diff_sigmai}. As $\sigma^*_i$ decreases, the maximum stress of the composite is strongly reduced and the transition point from the undamaged regime to this softening regime shifts to lower strains, while the slope of the softened stress-strain curve remains nearly unchanged. On the other hand, within the present model, the strain at break is only weakly affected by $\sigma_i^*$, with a slight tendency to increase as the interface becomes weaker.

\begin{figure}[htpb]
    \centering
    \includegraphics[width=\linewidth]{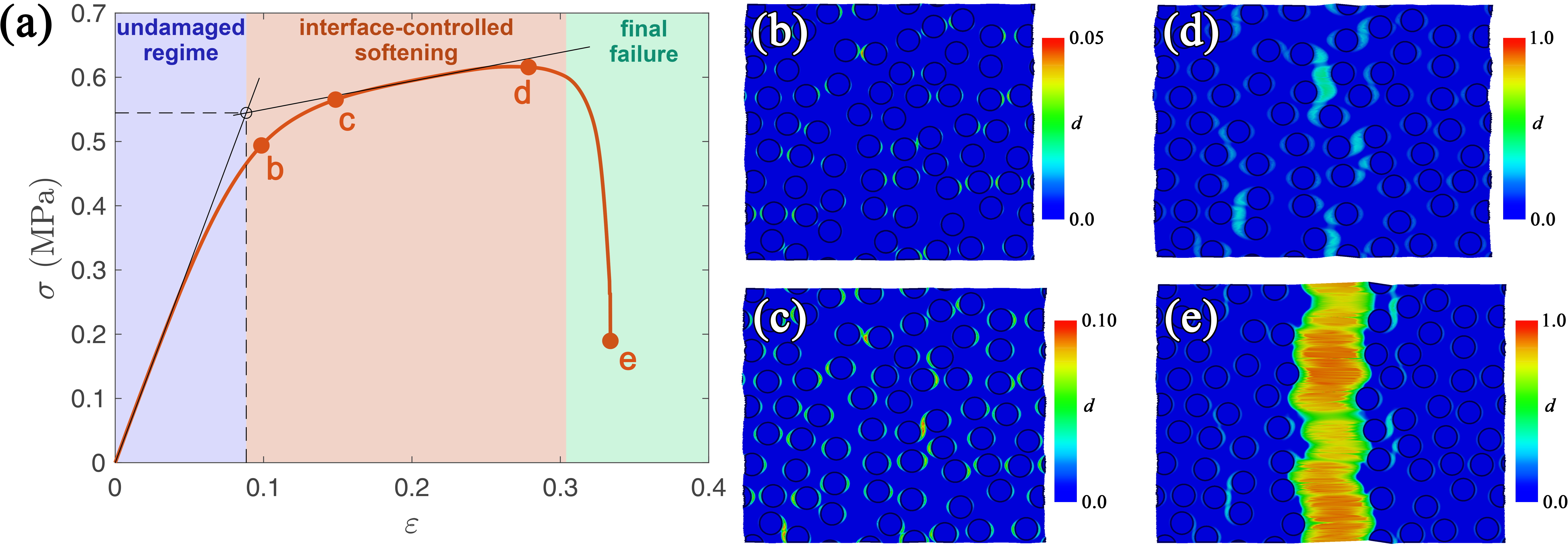}
    \caption{Multi-stage fracture of the polymer composite with weak interfaces $\sigma^*_i/\sigma^*_m = 0.33$. (a) Simulated stress-strain curve. Background colors indicate the undamaged regime (blue), interface-controlled softening (orange), and final failure (green). Black solid lines denote linear fit to the initial and intermediate regimes. The points marked b--e correspond to the crack phase-field snapshots shown in (b)--(e). (b--e) Distribution of the crack phase-field variable $d$ at (b) $\varepsilon = 0.10$; (c) $\varepsilon = 0.15$; (d) $\varepsilon = 0.28$; (e) $\varepsilon = 0.33$.}
    \label{fig:threeSectionFracture}
\end{figure}

This type of intermediate softening regime has been reported in many experimental studies on particle-filled  polymer composites, and is commonly attributed to progressive interfacial debonding\cite{Vollenberg1988,Asp1997,Renner2005}. To analyze the softening process in more detail within the present model, the case with $\sigma^*_i/\sigma^*_m = 0.33$ is selected as a representative example, and the corresponding stress-strain curve is depicted in Fig.~\ref{fig:threeSectionFracture}(a), where the macroscopic stress-strain curve is divided into three regimes. In the first undamaged regime, the influence of damage is negligible and does not affect the macroscopic response, so that the macroscopic stress-strain curve is the same as the undamaged nonlinear elastic behavior of the composite. With the further loading, the curve enters an interface-controlled softening regime, and the crack phase field at the beginning of damage is depicted in Fig.~\ref{fig:threeSectionFracture}(b-c). The damage appears mainly as crescent-shaped zones around the two tensile poles of the particles along the loading direction, representing interfacial debonding observed in microscopic experiments\cite{Bai2003}. Meanwhile, some neighboring damage zones start to merge locally, but the damage pattern remains spatially distributed rather than developing into a dominant crack band. This damage pattern reduces the load-carrying capacity of the affected regions, which is consistent with the reduced and nearly constant tangent stiffness observed in the intermediate softening regime. When the deformation is further increased, the curve reaches an abrupt turning point and the stress drops sharply. The crack phase fields near final failure shown in Fig.~\ref{fig:threeSectionFracture}(d-e) reveal that the distributed crack bands coalesce and evolve into a dominant crack band. As this crack band penetrates through the polymer matrix, a continuous fracture path is formed and the composite loses its macroscopic load-carrying capacity.

\begin{figure}[htpb]
    \centering
    \includegraphics[width=\linewidth]{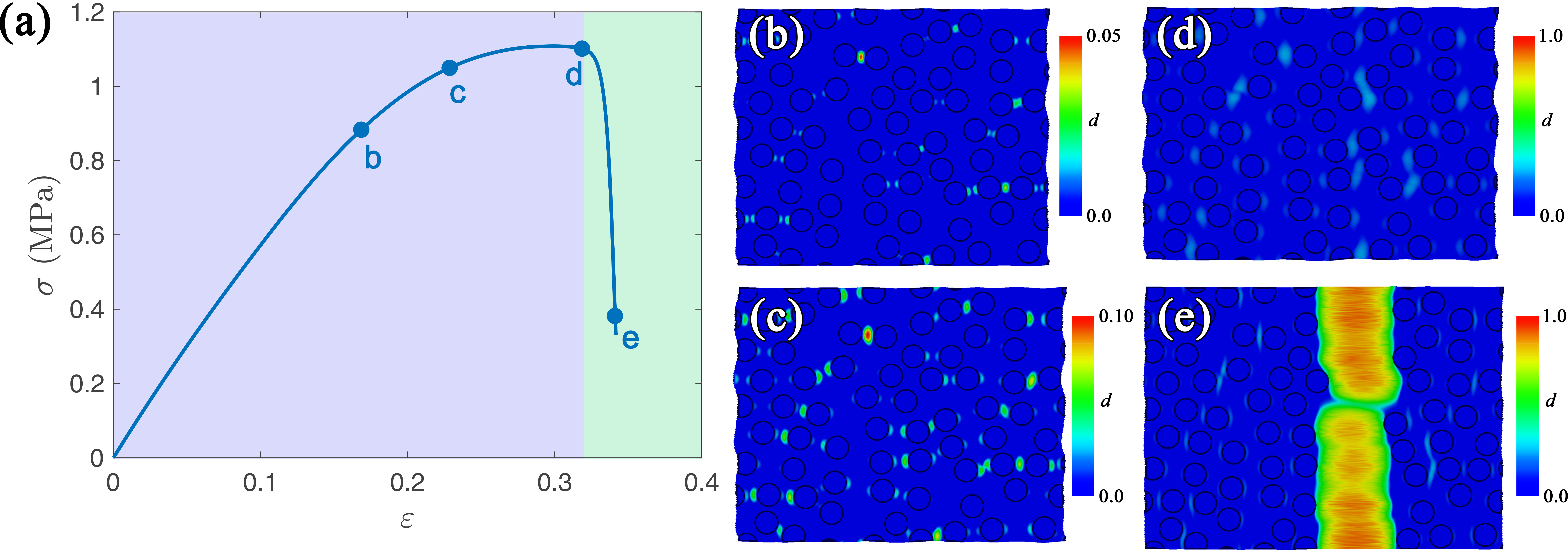}
    \caption{Two-stage fracture of the polymer composite with well-bonded interfaces $\sigma^*_i/\sigma^*_m = 1$ and $G_i/G_m = 1$. (a) Simulated stress-strain curve. Background colors indicate the undamaged regime (blue) and final failure (green). The points marked b--e correspond to the crack phase-field snapshots shown in (b)--(e). (b--e) Distribution of the crack phase-field variable $d$ at (b) $\varepsilon = 0.17$; (c) $\varepsilon = 0.23$; (d) $\varepsilon = 0.32$; (e) $\varepsilon = 0.34$.}
    \label{fig:twoSectionFracture}
\end{figure}

For comparison, a corresponding well-bonded reference case is considered, in which $\sigma^*_i/\sigma^*_m=1$ and $G_i/G_m=1$. The corresponding stress-strain curve and fracture process are shown in Fig.~\ref{fig:twoSectionFracture}. In this case, the interface has the same critical strength and fracture energy as the matrix and therefore does not act as a preferential damage-initiation site. As shown in Fig.~\ref{fig:twoSectionFracture}(a), the response first exhibits a nearly undamaged nonlinear elastic regime, during which the tangent stiffness remains relatively high and decreases gradually only when approaching the peak stress. The crack phase fields in Fig.~\ref{fig:twoSectionFracture}(b-c) show that damage nucleates in the narrow matrix regions between closely spaced particles, rather than along the particle-matrix interface. This damage-initiation pattern is consistent with experimental observations in well-bonded particle-filled polymer composites, where cavitation often initiates in the matrix near the particle poles\cite{Poulain2017}. After damage initiation, the curve quickly enters the final failure regime, characterized by an abrupt stress drop from the peak stress. Unlike the weak-interface case, no stable intermediate interface-controlled softening regime appears. Correspondingly, the localized matrix damage rapidly coalesces and concentrates into a dominant crack band across the RVE, leading to macroscopic failure of the polymer composite.

\subsection{Influence of Matrix and Interfacial Critical Energy Release Rates}

\begin{figure}[htpb]
    \centering
    \includegraphics[width=1.0\linewidth]{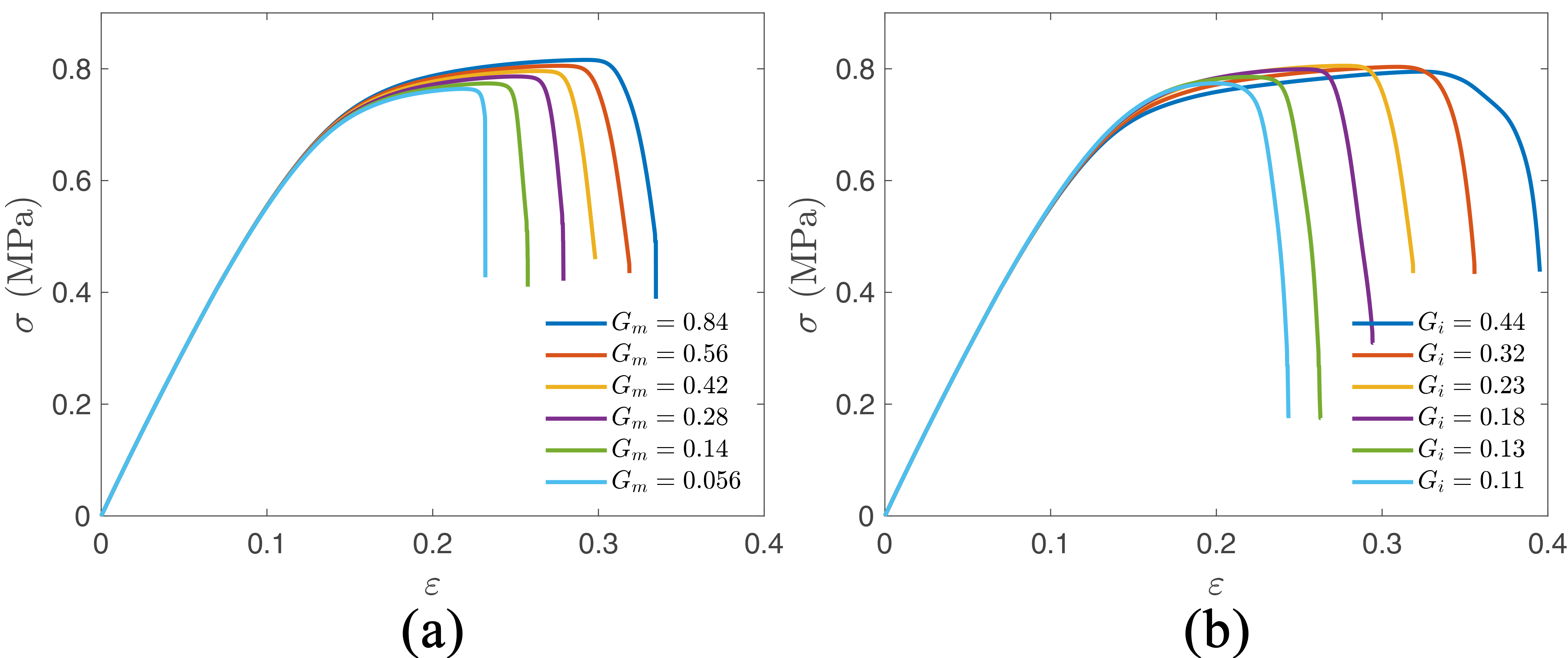}
    \caption{Effect of matrix and interfacial critical energy release rate on the macroscopic stress-strain response of the composite. (a) The matrix critical energy release rate $G_m$ varies from $0.056\ {\rm N/mm}$ to $0.84\ {\rm N/mm}$ at a fixed interfacial value $G_i = 0.23\ {\rm N/mm}$. (b) The interfacial critical energy release rate $G_i$ varies from $0.11\ {\rm N/mm}$ to $0.44\ {\rm N/mm}$ at a fixed matrix value $G_m = 0.56\ {\rm N/mm}$.}
    \label{fig:diffGmandGi}
\end{figure}

In addition to critical strength, the critical energy release rates also play an important role in the fracture process of the composite. We therefore further studied the mechanical response for different combinations of $G_m$ and $G_i$. In all cases, the polymer matrix strength is kept the same as in the previous cases, and the interface remains weaker in strength with $\sigma^*_i/\sigma^*_m = 0.56$. In Fig.~\ref{fig:diffGmandGi}(a), the interfacial critical energy release rate is fixed at $G_i = 0.23\ {\rm N/mm}$, while the matrix value $G_m$ is varied from $0.056\ {\rm N/mm}$ to $0.84\ {\rm N/mm}$. The results show that increasing $G_m$ leads to a larger strain at break and higher macroscopic toughness, while the transition from the initial undamaged regime to the interface-controlled softening regime is almost unaffected.

On the other hand, the corresponding study of $G_i$ is shown in Fig.~\ref{fig:diffGmandGi}(b), where the matrix critical energy release rate is fixed at $G_m = 0.56\ {\rm N/mm}$, and the interfacial value $G_i$ is varied from $0.11\ {\rm N/mm}$ to $0.44\ {\rm N/mm}$. A similar trend is observed: the strain at break increases with $G_i$, whereas the transition point of the softening regime remains nearly unchanged. This indicates that the transition is governed mainly by the interfacial strength, rather than by $G_m$ or $G_i$, over the investigated parameter range. Comparing Fig.~\ref{fig:diffGmandGi}(a) and (b), varying $G_i$ modifies the softening regime of the stress-strain curves more strongly than varying $G_m$ over the ranges considered. This behavior indicates that, in composites whose softening is controlled mainly by weak interfaces, the macroscopic elongation and energy dissipation are more sensitive to the interfacial fracture energy than to the matrix fracture energy, so that increasing $G_i$ may be a more effective way to improve the overall elongation and toughness of composites with weak interfaces.

Fig.~\ref{fig:diffGmandGi}(b) also includes the extreme case with a very low interfacial fracture energy with $G_i = 0.11\ {\rm N/mm}$. In this case, the stress-strain curve does not exhibit a clear interface-controlled softening regime. Instead, the stress increases continuously and failure occurs shortly after damage initiation. Damage still nucleates at the interface, but the very low $G_i$ allows the interface to dissipate only a small amount of elastic energy, leading to a rapid loss of load-carrying capacity in the interfacial region. The load is then transferred rapidly to the polymer matrix, causing the crack to propagate from the interface into the bulk matrix. As a result, the macroscopic mechanical response passes quickly from damage initiation to unstable crack path localization and final failure, without a stable interface-controlled softening regime.

Overall, the simulations show that when the interface is weaker in strength than the matrix, fracture initiates at the interface, and the strain of transition from undamaged regime to softening regime is mainly controlled by the interfacial strength, with only a weak influence from $G_m$ or $G_i$. In the present simulations, a distinct intermediate softening regime appears only in weak-interface systems with sufficient fracture energy to support progressive debonding throughout the whole domain. If the interface is not the weak point in the composite, or if $G_i$ is very small, this intermediate regime will no longer appear, and the macroscopic mechanical response rapidly proceeds from damage initiation to abrupt final failure.

In experiments, the appearance of a clear softening regime between the initial response and abrupt final failure may reflect progressive interfacial debonding caused by the weak interface. For rubber-like polymer composites, where matrix-related yielding is not expected to produce a similar softening response, such an intermediate regime can provide useful evidence for identifying progressive interfacial debonding and interpreting the associated damage-evolution process. Based on the present model, the stress at this transition is governed mainly by the interfacial strength, and increasing the interfacial strength can therefore be an effective way to raise the maximum stress of the composite. Nevertheless, the absence of a clear softening regime does not necessarily indicate a well-bonded interface, because a similar two-stage response may also occur when progressive debonding becomes unstable owing to insufficient fracture energy of the interface or matrix.

\section{Weak-Interface-Induced Enhancement of Macroscopic Elongation}

Weak interfacial adhesion usually promotes debonding and may reduce the effective toughness or strength of the polymer composite, but under certain conditions it can enhance macroscopic deformability or fracture resistance\cite{Thio2004}. In this section, the CZM-PF model is used to investigate the mechanism of weak interface increasing the macroscopic elongation of particle-filled polymer composites, especially the roles of interfacial strength and fracture energy.

\begin{figure}[htpb]
    \centering
    \includegraphics[width=0.8\linewidth]{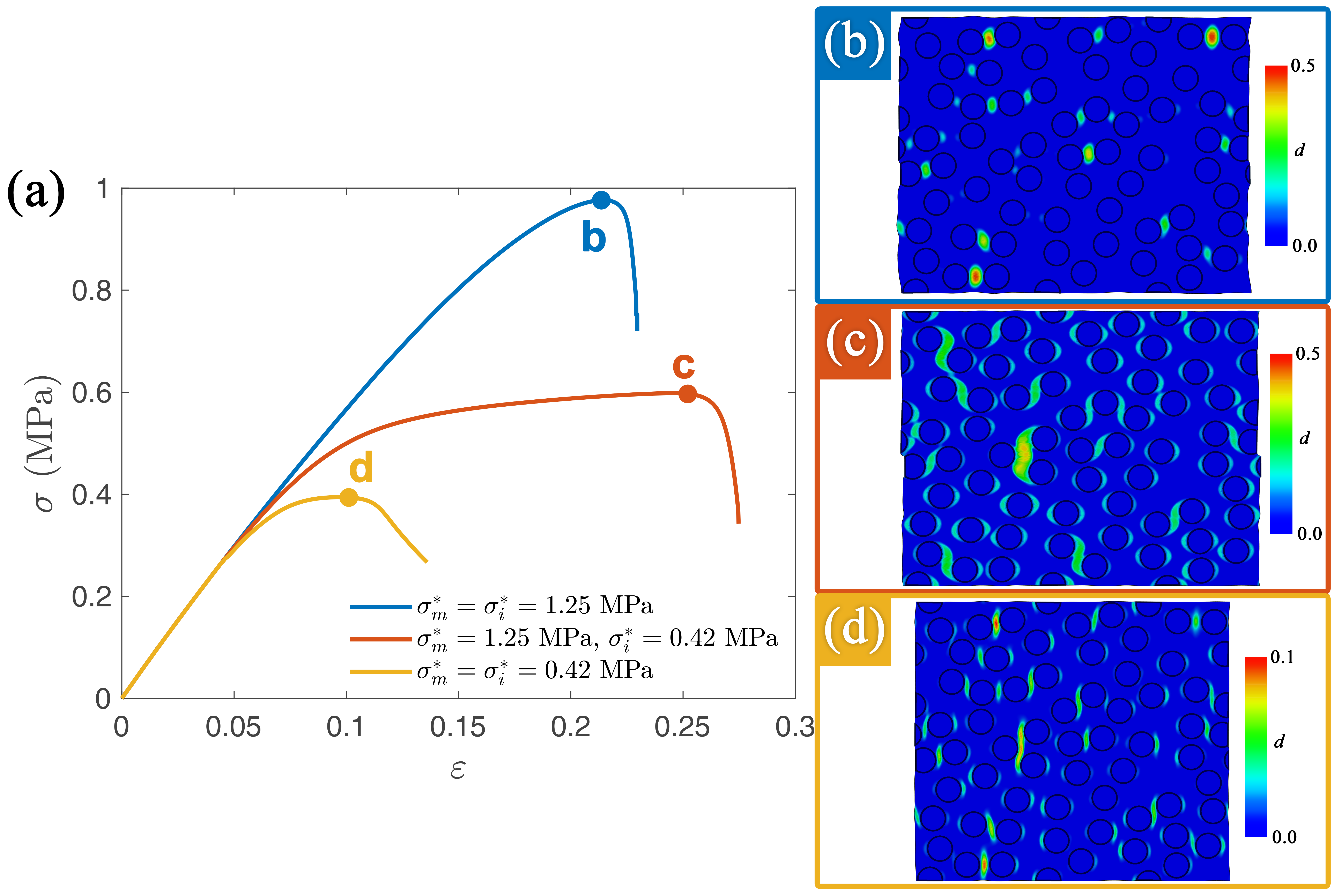}
    \caption{Comparison of three systems, including a low-strength well-bonded system ($\sigma^*_m = \sigma^*_i = 0.42\ {\rm MPa}$), a high-strength well-bonded system ($\sigma^*_m = \sigma^*_i = 1.25\ {\rm MPa}$), and a weak-interface system ($\sigma^*_m = 1.25\ {\rm MPa}$, $\sigma^*_i = 0.42\ {\rm MPa}$). (a) Macroscopic stress-strain curves for the three systems. The points marked b-d correspond to the crack phase-field snapshots shown in (b)-(d). (b--d) Distribution of the crack phase-field variable $d$ for (b) the high-strength well-bonded system, (c) the weak-interface system, and (d) the low-strength well-bonded system.}
    \label{fig:unsymInterface}
\end{figure}

To illustrate this mechanism, three systems are considered. The filler volume fraction is fixed at $f=0.42$, and the elastic properties of the polymer matrix and filler particles are kept identical to those used in the preceding analyses. The critical energy release rates of the matrix and interface are both set to $G_m=G_i=0.14\ {\rm N/mm}$. Two well-bonded reference systems are first considered, with high-strength reference $\sigma^*_m=\sigma^*_i=1.25\ {\rm MPa}$ and low-strength reference $\sigma^*_m=\sigma^*_i=0.42\ {\rm MPa}$. A weak-interface system is then introduced by setting $\sigma^*_m=1.25\ {\rm MPa}$ and $\sigma^*_i=0.42\ {\rm MPa}$ while keeping $G_m=G_i$. In this case, the lower interfacial strength promotes damage initiation at the particle--matrix interface. Comparing this weak-interface system with the two well-bonded reference systems allows us to distinguish the effect of selective interfacial strength reduction from that of an overall strength reduction. The corresponding stress-strain curves and damage patterns are presented in Fig.~\ref{fig:unsymInterface}.

As shown by the stress-strain curves in Fig.~\ref{fig:unsymInterface}(a), the low-strength reference system with $\sigma^*_m=\sigma^*_i=0.42\ {\rm MPa}$ exhibits the lowest maximum stress and the smallest strain at break, whereas the high-strength reference system with $\sigma^*_m=\sigma^*_i=1.25\ {\rm MPa}$ exhibits the highest maximum stress. Both well-bonded reference systems show a two-stage response consisting of an initial undamaged regime followed by a gradual decrease of stiffness to final fracture. For the weak-interface system with $\sigma^*_m=1.25\ {\rm MPa}$ and $\sigma^*_i=0.42\ {\rm MPa}$, the maximum stress lies between those of the two reference systems, but the strain at break is the largest among the three cases. In addition, a distinct interface-controlled softening regime with a nearly linear stress-strain response appears before final fracture.

To clarify the mechanism by which weak interfaces enhance elongation, the crack phase-field distributions just before final fracture are compared in Fig.~\ref{fig:unsymInterface}(b-d). The two well-bonded reference systems in Fig.~\ref{fig:unsymInterface}(b) and (d) exhibit similar damage patterns, with damage concentrated in a limited number of preferential regions, such as the narrow matrix ligaments between closely spaced particles and the matrix near the tensile poles of the particles. Since interfacial debonding does not occur in these systems, damage is activated mainly in the highly stressed matrix regions. These damaged regions can rapidly interact and coalesce through the continuous matrix, promoting the early formation of a dominant crack band. In contrast, the weak-interface system in Fig.~\ref{fig:unsymInterface}(c) develops crescent-shaped damage zones around nearly all particles, so that a larger portion of the composite is involved in the damage process. The weak interfaces therefore act as distributed weak sites throughout the composite, allowing interfacial debonding and the associated energy dissipation to spread more widely. This distributed debonding also relaxes local stress concentration around stiff particles by reducing particle-matrix load transfer, thereby weakening the tendency for damage to concentrate into a dominant crack band at an early stage. As a result, compared with the high-strength well-bonded reference systems, the weak-interface system exhibits a lower maximum stress, but fracture localization is delayed and the strain at break is increased.

\begin{figure}[htpb]
    \centering
    \includegraphics[width=1.0\textwidth]{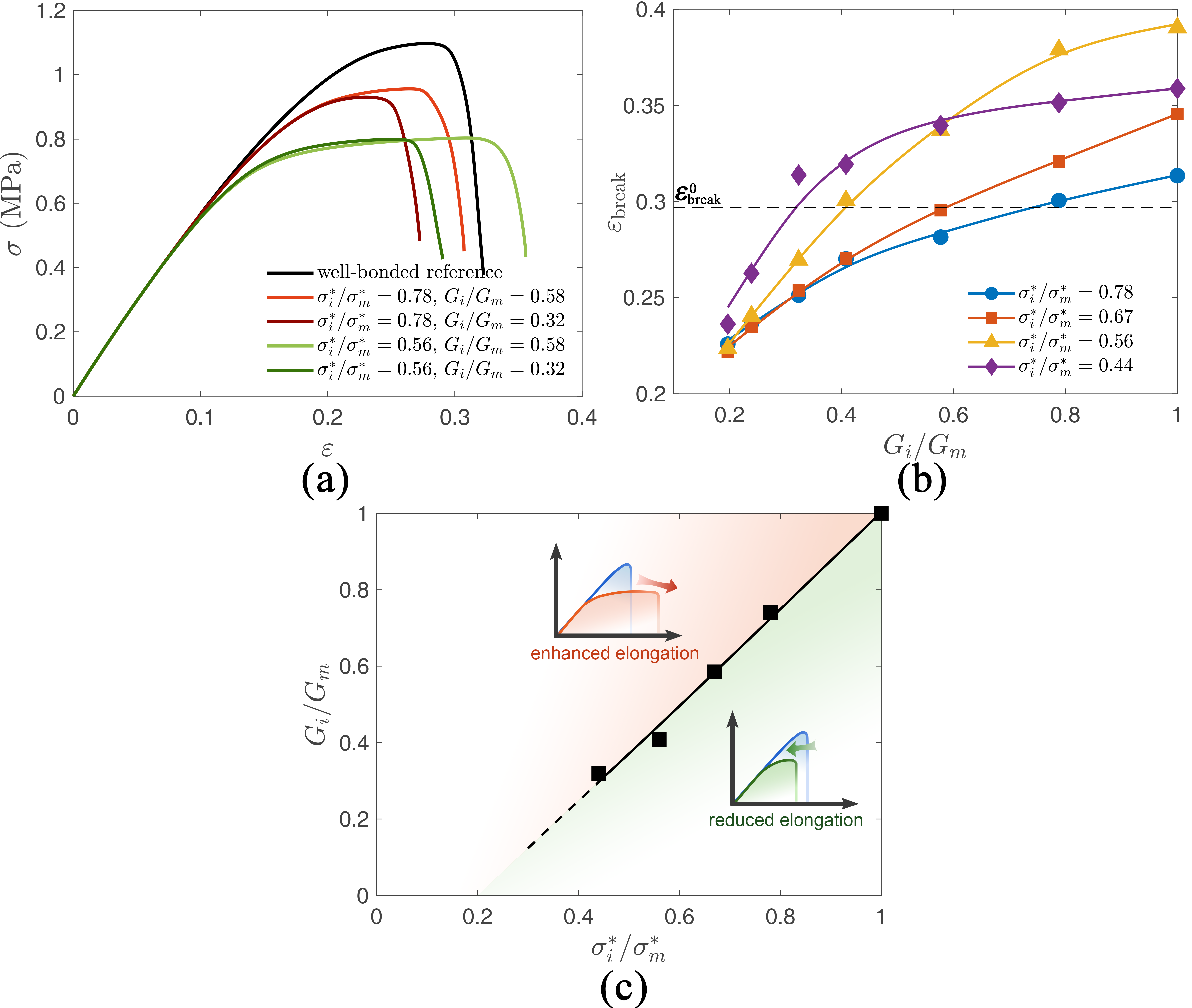}
    \caption{Combined effects of $\sigma^*_i$ and $G_i$ on the macroscopic elongation of the composite. (a) Stress-strain curves of the well-bonded reference system and representative weak-interface systems with different $\sigma^*_i/\sigma^*_m$ and $G_i/G_m$. (b) Nominal strain at break $\varepsilon_{\rm break}$ as a function of $G_i/G_m$ for different interfacial strength ratios. The horizontal black dashed line denotes the strain at break of the well-bonded reference system $\varepsilon_{\rm break}^0$. (c) Critical combinations of $\sigma^*_i/\sigma^*_m$ and $G_i/G_m$ for which $\varepsilon_{\rm break}=\varepsilon_{\rm break}^0$. Black symbols denote the critical points identified from the trends in (b), and the solid line represents the approximate transition boundary. The region above the boundary corresponds to enhanced elongation relative to the well-bonded reference, whereas the region below the boundary corresponds to reduced elongation.}
    \label{fig:analyze_epsbrk}
\end{figure}

However, the enhancement of macroscopic elongation by a weak interface is not guaranteed, but depends on the combined effects of interfacial strength and fracture energy. To demonstrate this dependence, $\sigma^*_i$ and $G_i$ are varied while the matrix parameters are fixed at $\sigma^*_m=1.25\ {\rm MPa}$ and $G_m=0.56\ {\rm N/mm}$. Fig.~\ref{fig:analyze_epsbrk}(a) compares the stress--strain curve of the well-bonded reference system ($\sigma^*_i=\sigma^*_m$ and $G_i=G_m$) with those of several weak-interface systems with different values of $\sigma^*_i$ and $G_i$. In all weak-interface cases, the lower interfacial strength reduces the maximum stress relative to the well-bonded reference, but the strain at break relative to the well-bonded system depends on $G_i$. For $\sigma^*_i/\sigma^*_m=0.56$, the case with $G_i/G_m=0.58$ exhibits a larger strain at break than the well-bonded system, whereas reducing the fracture-energy ratio to $G_i/G_m=0.32$ results in a smaller strain at break. Thus, introducing a weak interface can either enhance or reduce the macroscopic elongation, depending on whether the interface can gradually dissipate energy to sustain progressive debonding before final localization. When the interfacial strength ratio is increased to $\sigma^*_i/\sigma^*_m=0.78$, neither of the two considered $G_i/G_m$ values gives a strain at break greater than that of the well-bonded reference. These comparisons indicate that the interfacial fracture energy required for elongation enhancement depends on the interfacial strength, and that a higher $\sigma^*_i/\sigma^*_m$ generally requires a higher $G_i/G_m$ for the strain at break to exceed that of the well-bonded system.

To quantify the combined effects of interfacial strength and fracture energy, simulations are performed over the ranges from 0.20 to 1.00 for $G_i/G_m$ and from 0.44 to 0.78 for $\sigma^*_i/\sigma^*_m$. The nominal strain at break, $\varepsilon_{\rm break}$, extracted from these simulations is shown in Fig.~\ref{fig:analyze_epsbrk}(b). For each interfacial strength ratio, $\varepsilon_{\rm break}$ increases continuously with $G_i/G_m$. At a given fracture-energy ratio, a lower $\sigma^*_i/\sigma^*_m$ usually results in a larger strain at break. At fixed $G_i$, reducing $\sigma^*_i$ spreads the same amount of interfacial energy dissipation over a larger deformation range, so the interfacial load-carrying capacity decreases more gradually, which helps prevent premature damage transfer into the matrix, thereby delaying crack localization and increasing the strain at break.

The horizontal dashed line in Fig.~\ref{fig:analyze_epsbrk}(b) denotes the strain at break of the well-bonded reference system, $\varepsilon_{\rm break}^0$. For each $\sigma^*_i/\sigma^*_m$, the critical value of $G_i/G_m$ at which the smoothed $\varepsilon_{\rm break}$ trend reaches $\varepsilon_{\rm break}^0$ is identified, and these critical values are plotted as black symbols in Fig.~\ref{fig:analyze_epsbrk}(c). Over the investigated parameter range, the critical points form a monotonic boundary with a nearly linear trend. The region above this boundary corresponds to $\varepsilon_{\rm break}>\varepsilon_{\rm break}^0$, indicating that introducing a weak interface enhances the macroscopic elongation, whereas the region below the boundary corresponds to a reduction in strain at break. As the interfacial strength decreases, the minimum $G_i$ required for the strain at break to exceed the well-bonded reference also decreases.

The results therefore show that elongation enhancement is governed by the balance between interfacial strength and fracture energy. A reduction in interfacial strength promotes progressive debonding, and if the interfacial fracture energy remains sufficient to prevent a rapid loss of load-carrying capacity, this distributed debonding damage can delay localization into a dominant crack band and increase the macroscopic elongation. By contrast, if $G_i$ is too low, the interface rapidly loses its load-carrying capacity after damage initiation, leading to early crack localization and a lower strain at break. Although the precise location of the transition boundary depends on the composite system and parameter range considered, it captures the underlying competition between interfacial strength and fracture energy in determining whether weak interface enhances or reduces macroscopic elongation.

The present results provide a possible explanation for why weak interfaces can either enhance or reduce the macroscopic fracture performance observed in experiments. For example, Thio et al.\cite{Thio2004} reported that changing the surface treatment from CH to CF in glass-particle-filled polypropylene composite reduced both the interfacial strength and the work of adhesion. However, the reduction in interfacial strength was much more pronounced, whereas the work of adhesion decreased only slightly. Despite the weakening of both interfacial properties, the CF-treated composite exhibited a larger fracture displacement and higher toughness in single-edge-notched tests. This behavior can be explained by the present finding that the macroscopic effect of interfacial weakening is governed by the balance between interfacial strength and energy-dissipation capacity. A weak interface may increase the strain at break when the loss of interfacial fracture energy is limited relative to the reduction in strength; otherwise, it may lead to early failure.

\section{Conclusion}

This work investigates how weak particle-matrix interfaces affect the macroscopic tensile response and fracture of particle-filled polymer composites. To describe the coupled evolution and competition between interfacial debonding and matrix fracture, a finite-deformation framework was established by integrating a CZM phase-field formulation, a diffuse representation of the particle-matrix interface, and a hyperelastic description of the polymer matrix. Within this framework, interfacial debonding and matrix fracture are described by a common crack phase field, enabling their competing evolution to be examined consistently at the mesoscale.

The model was calibrated against experimental stress-strain curves of particle-filled polyurethane composites with different filler volume fractions and reproduced the main features of the tensile response and their variation with filler content, although the strain at break was quantitatively underestimated. The influence of filler volume fraction on the maximum stress was found to depend on the interfacial property. Increasing the filler volume fraction reduced the maximum stress in the weak-interface system but slightly increased in the well-bonded systems.

Weak interfacial strength was found to be able to produce a distinct multi-stage fracture process. When the interfacial strength is lower than that of the matrix, the stress-strain curve develops an intermediate softening regime with a reduced but nearly constant tangent stiffness. In this intermediate regime, crescent-shaped interfacial debonding develop around particles throughout the whole domain before eventually localizing into a dominant crack band. The interfacial strength mainly controls the onset of softening and the maximum stress, whereas the interfacial and matrix fracture energies mainly govern the subsequent damage evolution and strain at break. Therefore, a stable intermediate softening regime requires not only a weak interface in strength but also sufficient interfacial fracture energy to sustain progressive debonding. Very low interfacial fracture energy can lead to a rapid loss of interfacial load-carrying capacity and preventing the development of a distinct softening regime.

More importantly, weak interfaces can either increase or decrease the macroscopic strain at break. Compared with the well-bonded reference systems, the distributed debonding damage pattern introduced by weak interface can effectively dissipate energy and relax local stress concentration, thereby delaying damage localization and potentially increasing the strain at break. Parametric studies using the ratios $\sigma^*_i/\sigma^*_m$ and $G_i/G_m$ revealed a transition boundary between increased and reduced strain at break relative to the well-bonded reference. This transition reflects the combined roles of interfacial strength and fracture energy: reduced interfacial strength promotes distributed debonding, whereas sufficient interfacial fracture energy is also required to prevent rapid loss of interfacial load-carrying capacity. The macroscopic effect of weak interfaces is therefore governed by the balance between interfacial strength and fracture energy, rather than by either parameter alone.

Overall, the present study establishes a mechanistic link between interfacial fracture properties, mesoscale damage evolution, and macroscopic tensile failure in particle-filled polymer composites. This relationship explains the different experimental trends observed in polymer composites and provides a basis for evaluating interface-controlled mechanical performance. Nevertheless, the present framework does not account for temperature- and strain-rate-dependent behavior, which should be considered in the future to extend the study to viscoelastic polymer composites.




\section*{Declaration of Competing Interest}

The authors declare that they have no known competing financial interests or personal relationships that could have appeared to influence the work reported in this paper.

\section*{CRediT Authorship Contribution Statement}

Xu Chen: Conceptualization, Methodology, Software, Investigation, Writing - original draft. 
Ya Duan: Methodology, Software. 
Xiaoying Zhuang: Methodology, Writing - review \& editing. 
Timon Rabczuk: Conceptualization, Methodology, Writing - review \& editing. 

\section*{Data Availability}

Data are available on request.

\bibliographystyle{elsarticle-num} 
\bibliography{references}

@article{Zhen2024,
  title = {Phase-field modelling of fracture in viscoelastic composite using isogeometric {FCM}},
  volume = {274},
  url = {http://dx.doi.org/10.1016/j.ijmecsci.2024.109266},
  DOI = {10.1016/j.ijmecsci.2024.109266},
  journal = {International Journal of Mechanical Sciences},
  author = {Zhen,  Hao and Hu,  Pengmin and Liu,  Xiangyang and Dong,  Chunying},
  year = {2024},
  pages = {109266}
}

@article{Wu2017,
  title = {A unified phase-field theory for the mechanics of damage and quasi-brittle failure},
  volume = {103},
  url = {http://dx.doi.org/10.1016/j.jmps.2017.03.015},
  DOI = {10.1016/j.jmps.2017.03.015},
  journal = {Journal of the Mechanics and Physics of Solids},
  author = {Wu,  Jian-Ying},
  year = {2017},
  pages = {72--99}
}

@article{Wu2018,
  title = {A length scale insensitive phase-field damage model for brittle fracture},
  volume = {119},
  url = {http://dx.doi.org/10.1016/j.jmps.2018.06.006},
  DOI = {10.1016/j.jmps.2018.06.006},
  journal = {Journal of the Mechanics and Physics of Solids},
  author = {Wu,  Jian-Ying and Nguyen,  Vinh Phu},
  year = {2018},
  pages = {20--42}
}

@article{Mandal2020,
  title = {A length scale insensitive phase field model for brittle fracture of hyperelastic solids},
  volume = {236},
  url = {http://dx.doi.org/10.1016/j.engfracmech.2020.107196},
  DOI = {10.1016/j.engfracmech.2020.107196},
  journal = {Engineering Fracture Mechanics},
  author = {Mandal,  Tushar Kanti and Gupta,  Abhinav and Nguyen,  Vinh Phu and Chowdhury,  Rajib and de Vaucorbeil, Alban},
  year = {2020},
  pages = {107196}
}

@article{HansenDrr2020,
  title = {Phase-field modeling of crack branching and deflection in heterogeneous media},
  volume = {232},
  url = {http://dx.doi.org/10.1016/j.engfracmech.2020.107004},
  DOI = {10.1016/j.engfracmech.2020.107004},
  journal = {Engineering Fracture Mechanics},
  author = {Hansen-D\"{o}rr,  Arne Claus and Dammaß,  Franz and de Borst,  René and K\"{a}stner,  Markus},
  year = {2020},
  pages = {107004}
}

@article{Molnar2017,
  title = {{2D} and {3D} {Abaqus} implementation of a robust staggered phase-field solution for modeling brittle fracture},
  volume = {130},
  url = {http://dx.doi.org/10.1016/j.finel.2017.03.002},
  DOI = {10.1016/j.finel.2017.03.002},
  journal = {Finite Elements in Analysis and Design},
  author = {Molnár,  Gergely and Gravouil,  Anthony},
  year = {2017},
  pages = {27--38}
}

@article{Nguyen2014,
  title = {Discontinuous {Galerkin}/extrinsic cohesive zone modeling: Implementation caveats and applications in computational fracture mechanics},
  volume = {128},
  url = {http://dx.doi.org/10.1016/j.engfracmech.2014.07.003},
  DOI = {10.1016/j.engfracmech.2014.07.003},
  journal = {Engineering Fracture Mechanics},
  author = {Nguyen,  Vinh Phu},
  year = {2014},
  pages = {37--68}
}

@article{HansenDrr2019,
  title = {Phase-field modelling of interface failure in brittle materials},
  volume = {346},
  url = {http://dx.doi.org/10.1016/j.cma.2018.11.020},
  DOI = {10.1016/j.cma.2018.11.020},
  journal = {Computer Methods in Applied Mechanics and Engineering},
  author = {Hansen-D\"{o}rr,  Arne Claus and de Borst,  René and Hennig,  Paul and K\"{a}stner,  Markus},
  year = {2019},
  pages = {25--42}
}

@article{Bai2003,
  title = {Interfacial debonding behavior of a rigid particle-filled polymer composite},
  volume = {10},
  url = {http://dx.doi.org/10.1163/156855403765826892},
  DOI = {10.1163/156855403765826892},
  number = {2--3},
  journal = {Composite Interfaces},
  author = {Bai,  Shu-Lin and Wang,  Min and Zhao,  Xue-Feng},
  year = {2003},
  pages = {243--253}
}

@article{Poulain2017,
  title = {Damage in elastomers: nucleation and growth of cavities,  micro-cracks,  and macro-cracks},
  volume = {205},
  url = {http://dx.doi.org/10.1007/s10704-016-0176-9},
  DOI = {10.1007/s10704-016-0176-9},
  number = {1},
  journal = {International Journal of Fracture},
  author = {Poulain,  X. and Lefèvre,  V. and Lopez-Pamies,  O. and Ravi-Chandar,  K.},
  year = {2017},
  pages = {1--21}
}

@article{Thio2004,
  title = {Role of interfacial adhesion strength on toughening polypropylene with rigid particles},
  volume = {45},
  url = {http://dx.doi.org/10.1016/j.polymer.2004.02.064},
  DOI = {10.1016/j.polymer.2004.02.064},
  number = {10},
  journal = {Polymer},
  author = {Thio,  Y.S. and Argon,  A.S. and Cohen,  R.E.},
  year = {2004},
  pages = {3139--3147}
}

@article{Francfort1998,
  title = {Revisiting brittle fracture as an energy minimization problem},
  volume = {46},
  url = {http://dx.doi.org/10.1016/S0022-5096(98)00034-9},
  DOI = {10.1016/s0022-5096(98)00034-9},
  number = {8},
  journal = {Journal of the Mechanics and Physics of Solids},
  author = {Francfort,  G.A. and Marigo,  J.-J.},
  year = {1998},
  pages = {1319--1342}
}

@article{Bourdin2000,
  title = {Numerical experiments in revisited brittle fracture},
  volume = {48},
  url = {http://dx.doi.org/10.1016/S0022-5096(99)00028-9},
  DOI = {10.1016/s0022-5096(99)00028-9},
  number = {4},
  journal = {Journal of the Mechanics and Physics of Solids},
  author = {Bourdin,  B. and Francfort,  G.A. and Marigo,  J.-J.},
  year = {2000},
  pages = {797--826}
}

@article{Miehe2010,
  title = {Thermodynamically consistent phase-field models of fracture: Variational principles and multi‐field {FE} implementations},
  volume = {83},
  url = {http://dx.doi.org/10.1002/nme.2861},
  DOI = {10.1002/nme.2861},
  number = {10},
  journal = {International Journal for Numerical Methods in Engineering},
  author = {Miehe,  C. and Welschinger,  F. and Hofacker,  M.},
  year = {2010},
  pages = {1273--1311}
}

@article{Nguyen2016,
  title = {A phase-field method for computational modeling of interfacial damage interacting with crack propagation in realistic microstructures obtained by microtomography},
  volume = {312},
  url = {http://dx.doi.org/10.1016/j.cma.2015.10.007},
  DOI = {10.1016/j.cma.2015.10.007},
  journal = {Computer Methods in Applied Mechanics and Engineering},
  author = {Nguyen,  T.T. and Yvonnet,  J. and Zhu,  Q.-Z. and Bornert,  M. and Chateau,  C.},
  year = {2016},
  pages = {567--595}
}

@article{Zhang2019,
  title = {Modelling progressive failure in multi-phase materials using a phase field method},
  volume = {209},
  url = {http://dx.doi.org/10.1016/j.engfracmech.2019.01.021},
  DOI = {10.1016/j.engfracmech.2019.01.021},
  journal = {Engineering Fracture Mechanics},
  author = {Zhang,  Peng and Hu,  Xiaofei and Yang,  Shangtong and Yao,  Weian},
  year = {2019},
  pages = {105--124}
}

@article{Miehe2014,
  title = {Phase field modeling of fracture in rubbery polymers. Part {I}: Finite elasticity coupled with brittle failure},
  volume = {65},
  url = {http://dx.doi.org/10.1016/j.jmps.2013.06.007},
  DOI = {10.1016/j.jmps.2013.06.007},
  journal = {Journal of the Mechanics and Physics of Solids},
  author = {Miehe,  Christian and Sch\"{a}nzel,  Lisa-Marie},
  year = {2014},
  pages = {93--113}
}

@article{Kundie2018,
  title = {Effects of Filler Size on the Mechanical Properties of Polymer-filled Dental Composites: A Review of Recent Developments},
  volume = {29},
  url = {http://dx.doi.org/10.21315/jps2018.29.1.10},
  DOI = {10.21315/jps2018.29.1.10},
  number = {1},
  journal = {Journal of Physical Science},
  author = {Kundie,  Fathie and Azhari,  Che Husna and Muchtar,  Andanastuti and Ahmad,  Zainal Arifin},
  year = {2018},
  pages = {141--165}
}

@article{Fu2008,
  title = {Effects of particle size,  particle/matrix interface adhesion and particle loading on mechanical properties of particulate--polymer composites},
  volume = {39},
  url = {http://dx.doi.org/10.1016/j.compositesb.2008.01.002},
  DOI = {10.1016/j.compositesb.2008.01.002},
  number = {6},
  journal = {Composites Part B: Engineering},
  author = {Fu,  Shao-Yun and Feng,  Xi-Qiao and Lauke,  Bernd and Mai,  Yiu-Wing},
  year = {2008},
  pages = {933--961}
}

@article{Fan2015,
  title = {Glass interface effect on high-strain-rate tensile response of a soft polyurethane elastomeric polymer material},
  volume = {118},
  url = {http://dx.doi.org/10.1016/j.compscitech.2015.08.007},
  DOI = {10.1016/j.compscitech.2015.08.007},
  journal = {Composites Science and Technology},
  author = {Fan,  J.T. and Weerheijm,  J. and Sluys,  L.J.},
  year = {2015},
  pages = {55--62}
}

@article{Aliotta2019,
  title = {Rigid filler toughening in {PLA}-Calcium Carbonate composites: Effect of particle surface treatment and matrix plasticization},
  volume = {113},
  url = {http://dx.doi.org/10.1016/j.eurpolymj.2018.12.042},
  DOI = {10.1016/j.eurpolymj.2018.12.042},
  journal = {European Polymer Journal},
  author = {Aliotta,  Laura and Cinelli,  Patrizia and Coltelli,  Maria Beatrice and Lazzeri,  Andrea},
  year = {2019},
  pages = {78--88}
}

@article{Zebarjad2004,
  title = {Influence of filler particles on deformation and fracture mechanism of isotactic polypropylene},
  volume = {155--156},
  url = {http://dx.doi.org/10.1016/j.jmatprotec.2004.04.187},
  DOI = {10.1016/j.jmatprotec.2004.04.187},
  journal = {Journal of Materials Processing Technology},
  author = {Zebarjad,  S.M. and Tahani,  M. and Sajjadi,  S.A.},
  year = {2004},
  pages = {1459--1464}
}

@inbook{Schwarzl1967,
  title = {On Mechanical Properties of Unfilled and Filled Elastomers},
  ISBN = {9781483198378},
  url = {http://dx.doi.org/10.1016/B978-1-4831-9837-8.50027-2},
  DOI = {10.1016/b978-1-4831-9837-8.50027-2},
  booktitle = {Mechanics and Chemistry of Solid Propellants},
  publisher = {Elsevier},
  author = {Schwarzl,  F.R. and Bree,  H.W. and Nederveen,  C.J. and Struik,  L.C.E. and Van der Wal,  C.W.},
  year = {1967},
  pages = {503--538}
}

@article{Nunes2000,
  title = {Polymer--filler interactions and mechanical properties of a polyurethane elastomer},
  volume = {19},
  url = {http://dx.doi.org/10.1016/S0142-9418(98)00075-0},
  DOI = {10.1016/s0142-9418(98)00075-0},
  number = {1},
  journal = {Polymer Testing},
  author = {Nunes,  R.C.R. and Fonseca,  J.L.C. and Pereira,  M.R.},
  year = {2000},
  pages = {93--103}
}

@inbook{Bommegowda2021,
  title = {Role of Fillers in Controlling the Properties of Polymer Composites: A Review},
  ISBN = {9783030699253},
  url = {http://dx.doi.org/10.1007/978-3-030-69925-3\_62},
  DOI = {10.1007/978-3-030-69925-3\_62},
  booktitle = {Techno-Societal 2020},
  publisher = {Springer International Publishing},
  author = {Bommegowda,  K. B. and Renukappa,  N. M. and Rajan,  J. Sundara},
  year = {2021},
  pages = {637--648}
}

@article{Tao2013,
  title = {Microstructure Deformation and Fracture Mechanism of Highly Filled Polymer Composites under Large Tensile Deformation},
  volume = {419},
  url = {http://dx.doi.org/10.1088/1742-6596/419/1/012014},
  DOI = {10.1088/1742-6596/419/1/012014},
  journal = {Journal of Physics: Conference Series},
  author = {Zhang, Jiang Tao and Song, Dan Ping and Zhang, Mei and Zhai, Peng Cheng},
  year = {2013},
  pages = {012014}
}

@article{Kashfipour2018,
  title = {A review on the role of interface in mechanical,  thermal,  and electrical properties of polymer composites},
  volume = {1},
  url = {http://dx.doi.org/10.1007/s42114-018-0022-9},
  DOI = {10.1007/s42114-018-0022-9},
  number = {3},
  journal = {Advanced Composites and Hybrid Materials},
  author = {Kashfipour,  Marjan Alsadat and Mehra,  Nitin and Zhu,  Jiahua},
  year = {2018},
  pages = {415--439}
}

@article{Gent1984,
  title = {Failure processes in elastomers at or near a rigid spherical inclusion},
  volume = {19},
  url = {http://dx.doi.org/10.1007/BF00550265},
  DOI = {10.1007/bf00550265},
  number = {6},
  journal = {Journal of Materials Science},
  author = {Gent,  A. N. and Park,  Byoungkyeu},
  year = {1984},
  pages = {1947--1956}
}

@article{Toulemonde2016,
  title = {On the account of a cohesive interface for modeling the behavior until break of highly filled elastomers},
  volume = {93},
  url = {http://dx.doi.org/10.1016/j.mechmat.2015.09.014},
  DOI = {10.1016/j.mechmat.2015.09.014},
  journal = {Mechanics of Materials},
  author = {Toulemonde,  Paul-Aymé and Diani,  Julie and Gilormini,  Pierre and Desgardin,  Nancy},
  year = {2016},
  pages = {124--133}
}

@article{LeGulluche2023,
  title = {Role of Polymer--Particle Adhesion in the Reinforcement of Hybrid Hydrogels},
  volume = {56},
  url = {http://dx.doi.org/10.1021/acs.macromol.3c00745},
  DOI = {10.1021/acs.macromol.3c00745},
  number = {19},
  journal = {Macromolecules},
  author = {Le Gulluche,  Anne-Charlotte and Pantoustier,  Nadège and Brûlet,  Annie and Sanseau,  Olivier and Sotta,  Paul and Marcellan,  Alba},
  year = {2023},
  pages = {8024--8038}
}

@article{Rong2006,
  title = {Surface modification of nanoscale fillers for improving properties of polymer nanocomposites: A review},
  volume = {22},
  url = {http://dx.doi.org/10.1179/174328406X101247},
  DOI = {10.1179/174328406x101247},
  number = {7},
  journal = {Materials Science and Technology},
  author = {Rong,  M. Z. and Zhang,  M. Q. and Ruan,  W. H.},
  year = {2006},
  pages = {787--796}
}

@article{Ippolito2020,
  title = {Influence of the Surface Modification of Calcium Carbonate on Polyamide 12 Composites},
  volume = {12},
  url = {http://dx.doi.org/10.3390/polym12061295},
  DOI = {10.3390/polym12061295},
  number = {6},
  journal = {Polymers},
  author = {Ippolito,  Fabio and H\"{u}bner,  Gunter and Claypole,  Tim and Gane,  Patrick},
  year = {2020},
  pages = {1295}
}

@article{Bi2022,
  title = {Effect of Silane Coupling Agents on the Rheology,  Dynamic and Mechanical Properties of Ethylene Propylene Diene Rubber/Calcium Carbonate Composites},
  volume = {14},
  url = {http://dx.doi.org/10.3390/polym14163393},
  DOI = {10.3390/polym14163393},
  number = {16},
  journal = {Polymers},
  author = {Bi,  Weina and Goegelein,  Christoph and Hoch,  Martin and Kirchhoff,  Joerg and Zhao,  Shugao},
  year = {2022},
  pages = {3393}
}

@article{Hsieh2010,
  title = {The mechanisms and mechanics of the toughening of epoxy polymers modified with silica nanoparticles},
  volume = {51},
  url = {http://dx.doi.org/10.1016/j.polymer.2010.10.048},
  DOI = {10.1016/j.polymer.2010.10.048},
  number = {26},
  journal = {Polymer},
  author = {Hsieh,  T.H. and Kinloch,  A.J. and Masania,  K. and Taylor,  A.C. and Sprenger,  S.},
  year = {2010},
  pages = {6284--6294}
}

@article{Quaresimin2016,
  title = {Toughening mechanisms in polymer nanocomposites: From experiments to modelling},
  volume = {123},
  url = {http://dx.doi.org/10.1016/j.compscitech.2015.11.027},
  DOI = {10.1016/j.compscitech.2015.11.027},
  journal = {Composites Science and Technology},
  author = {Quaresimin,  M. and Schulte,  K. and Zappalorto,  M. and Chandrasekaran,  S.},
  year = {2016},
  pages = {187--204}
}

@article{Vollenberg1988,
  title = {Experimental determination of thermal and adhesion stress in particle filled thermoplasts},
  volume = {9},
  url = {http://dx.doi.org/10.1002/pc.750090603},
  DOI = {10.1002/pc.750090603},
  number = {6},
  journal = {Polymer Composites},
  author = {Vollenberg,  Peter and Heikens,  D. and Ladan,  H. C. B.},
  year = {1988},
  pages = {382--388}
}

@article{Asp1997,
  title = {Prediction of failure initiation in polypropylene with glass beads},
  volume = {18},
  url = {http://dx.doi.org/10.1002/pc.10256},
  DOI = {10.1002/pc.10256},
  number = {1},
  journal = {Polymer Composites},
  author = {Asp,  L. E. and Sj\"{o}gren,  B. A. and Berglund,  L. A.},
  year = {1997},
  pages = {9--15}
}

@article{Renner2005,
  title = {Analysis of the debonding process in polypropylene model composites},
  volume = {41},
  url = {http://dx.doi.org/10.1016/j.eurpolymj.2005.05.025},
  DOI = {10.1016/j.eurpolymj.2005.05.025},
  number = {11},
  journal = {European Polymer Journal},
  author = {Renner,  Károly and Yang,  Min Soo and Móczó,  János and Choi,  Hyoung Jin and Pukánszky,  Béla},
  year = {2005},
  pages = {2520--2529}
}

@article{Gilormini2017,
  title = {Stress-strain response and volume change of a highly filled rubbery composite: Experimental measurements and numerical simulations},
  volume = {111},
  url = {http://dx.doi.org/10.1016/j.mechmat.2017.05.006},
  DOI = {10.1016/j.mechmat.2017.05.006},
  journal = {Mechanics of Materials},
  author = {Gilormini,  Pierre and Toulemonde,  Paul-Aymé and Diani,  Julie and Gardere,  Antoine},
  year = {2017},
  pages = {57--65}
}

@article{Kun2021,
  title = {The Role of Interfacial Adhesion in Polymer Composites Engineered from Lignocellulosic Agricultural Waste},
  volume = {13},
  url = {http://dx.doi.org/10.3390/polym13183099},
  DOI = {10.3390/polym13183099},
  number = {18},
  journal = {Polymers},
  author = {Kun,  Dávid and Kárpáti,  Zoltán and Fekete,  Erika and Móczó,  János},
  year = {2021},
  pages = {3099}
}

@article{Cheng2007,
  title = {Void interaction and coalescence in polymeric materials},
  volume = {44},
  url = {http://dx.doi.org/10.1016/j.ijsolstr.2006.08.007},
  DOI = {10.1016/j.ijsolstr.2006.08.007},
  number = {6},
  journal = {International Journal of Solids and Structures},
  author = {Cheng,  L. and Guo,  T.F.},
  year = {2007},
  pages = {1787--1808}
}

@article{Lai2024,
  title = {Interfacial debonding and cracking in a solid propellant composite under uniaxial tension: An in situ synchrotron {X-ray} tomography study},
  volume = {256},
  url = {http://dx.doi.org/10.1016/j.compscitech.2024.110743},
  DOI = {10.1016/j.compscitech.2024.110743},
  journal = {Composites Science and Technology},
  author = {Lai,  G.D. and Sang,  L.P. and Bian,  Y.L. and Xie,  H.L. and Liu,  J.H. and Chai,  H.W.},
  year = {2024},
  pages = {110743}
}

@article{deFrancqueville2020,
  title = {Relationship between local damage and macroscopic response of soft materials highly reinforced by monodispersed particles},
  volume = {146},
  url = {http://dx.doi.org/10.1016/j.mechmat.2020.103408},
  DOI = {10.1016/j.mechmat.2020.103408},
  journal = {Mechanics of Materials},
  author = {de Francqueville,  Foucault and Gilormini,  Pierre and Diani,  Julie and Vandenbroucke,  Aude},
  year = {2020},
  pages = {103408}
}

@article{Brassart2009,
  title = {An extended {Mori--Tanaka} homogenization scheme for finite strain modeling of debonding in particle-reinforced elastomers},
  volume = {45},
  url = {http://dx.doi.org/10.1016/j.commatsci.2008.06.021},
  DOI = {10.1016/j.commatsci.2008.06.021},
  number = {3},
  journal = {Computational Materials Science},
  author = {Brassart,  L. and Inglis,  H.M. and Delannay,  L. and Doghri,  I. and Geubelle,  P.H.},
  year = {2009},
  pages = {611--616}
}

@article{Firooz2021,
  title = {Homogenization of Composites With Extended General Interfaces: Comprehensive Review and Unified Modeling},
  volume = {73},
  url = {http://dx.doi.org/10.1115/1.4051481},
  DOI = {10.1115/1.4051481},
  number = {4},
  journal = {Applied Mechanics Reviews},
  author = {Firooz,  S. and Steinmann,  P. and Javili,  A.},
  year = {2021},
  pages = {040802}
}

@article{Liu2015,
  title = {Influence of nanoparticle surface treatment on particle dispersion and interfacial adhesion in low-density polyethylene/aluminium oxide nanocomposites},
  volume = {66},
  url = {http://dx.doi.org/10.1016/j.eurpolymj.2015.01.046},
  DOI = {10.1016/j.eurpolymj.2015.01.046},
  journal = {European Polymer Journal},
  author = {Liu,  D. and Pourrahimi,  A.M. and Olsson,  R.T. and Hedenqvist,  M.S. and Gedde,  U.W.},
  year = {2015},
  pages = {67--77}
}

@article{Tann2018,
  title = {Crack nucleation in variational phase-field models of brittle fracture},
  volume = {110},
  url = {http://dx.doi.org/10.1016/j.jmps.2017.09.006},
  DOI = {10.1016/j.jmps.2017.09.006},
  journal = {Journal of the Mechanics and Physics of Solids},
  author = {Tanné,  E. and Li,  T. and Bourdin,  B. and Marigo,  J.-J. and Maurini,  C.},
  year = {2018},
  pages = {80--99}
}

@article{Loew2019,
  title = {Rate-dependent phase-field damage modeling of rubber and its experimental parameter identification},
  volume = {127},
  url = {http://dx.doi.org/10.1016/j.jmps.2019.03.022},
  DOI = {10.1016/j.jmps.2019.03.022},
  journal = {Journal of the Mechanics and Physics of Solids},
  author = {Loew,  Pascal J. and Peters,  Bernhard and Beex,  Lars A.A.},
  year = {2019},
  pages = {266--294}
}

@article{Chen2025,
  title = {A length-scale insensitive cohesive phase-field interface model: Application to concurrent bulk and interface fracture simulation in {Lithium-ion} battery materials},
  volume = {196},
  url = {http://dx.doi.org/10.1016/j.jmps.2024.106013},
  DOI = {10.1016/j.jmps.2024.106013},
  journal = {Journal of the Mechanics and Physics of Solids},
  author = {Chen,  Wan-Xin and Peng,  Xiang-Long and Wu,  Jian-Ying and Furat,  Orkun and Schmidt,  Volker and Xu,  Bai-Xiang},
  year = {2025},
  pages = {106013}
}

@article{Ambati2015,
  title = {A review on phase-field models of brittle fracture and a new fast hybrid formulation},
  volume = {55},
  url = {http://dx.doi.org/10.1007/s00466-014-1109-y},
  DOI = {10.1007/s00466-014-1109-y},
  number = {2},
  journal = {Computational Mechanics},
  author = {Ambati,  Marreddy and Gerasimov,  Tymofiy and De Lorenzis,  Laura},
  year = {2015},
  pages = {383--405}
}

@article{Ren2024,
  title = {Variational damage model: A novel consistent approach to fracture},
  volume = {305},
  url = {http://dx.doi.org/10.1016/j.compstruc.2024.107518},
  DOI = {10.1016/j.compstruc.2024.107518},
  journal = {Computers \& Structures},
  author = {Ren,  Huilong and Zhuang,  Xiaoying and Zhu,  Hehua and Rabczuk,  Timon},
  year = {2024},
  pages = {107518}
}

@article{Duan2026,
  title = {A unified variational damage model and an efficient length scale insensitive phase-field model},
  volume = {208},
  url = {http://dx.doi.org/10.1016/j.jmps.2025.106494},
  DOI = {10.1016/j.jmps.2025.106494},
  journal = {Journal of the Mechanics and Physics of Solids},
  author = {Duan,  Ya and Ren,  Huilong and Bie,  Yehui and Zhuang,  Xiaoying and Rabczuk,  Timon},
  year = {2026},
  pages = {106494}
}

@article{Damma2023,
  title = {Phase-field modelling and analysis of rate-dependent fracture phenomena at finite deformation},
  volume = {72},
  url = {http://dx.doi.org/10.1007/s00466-023-02310-1},
  DOI = {10.1007/s00466-023-02310-1},
  number = {5},
  journal = {Computational Mechanics},
  author = {Dammaß,  Franz and Kalina,  Karl A. and Ambati,  Marreddy and K\"{a}stner,  Markus},
  year = {2023},
  pages = {859--883}
}

@article{Geelen2019,
  title = {A phase-field formulation for dynamic cohesive fracture},
  volume = {348},
  url = {http://dx.doi.org/10.1016/j.cma.2019.01.026},
  DOI = {10.1016/j.cma.2019.01.026},
  journal = {Computer Methods in Applied Mechanics and Engineering},
  author = {Geelen,  Rudy J.M. and Liu,  Yingjie and Hu,  Tianchen and Tupek,  Michael R. and Dolbow,  John E.},
  year = {2019},
  pages = {680--711}
}

@article{Talamini2018,
  title = {Progressive damage and rupture in polymers},
  volume = {111},
  url = {http://dx.doi.org/10.1016/j.jmps.2017.11.013},
  DOI = {10.1016/j.jmps.2017.11.013},
  journal = {Journal of the Mechanics and Physics of Solids},
  author = {Talamini,  Brandon and Mao,  Yunwei and Anand,  Lallit},
  year = {2018},
  pages = {434--457}
}

@article{Moraleda2009,
  title = {Effect of interface fracture on the tensile deformation of fiber-reinforced elastomers},
  volume = {46},
  url = {http://dx.doi.org/10.1016/j.ijsolstr.2009.08.020},
  DOI = {10.1016/j.ijsolstr.2009.08.020},
  number = {25-26},
  journal = {International Journal of Solids and Structures},
  author = {Moraleda,  Joaquín and Segurado,  Javier and Llorca,  Javier},
  year = {2009},
  pages = {4287--4297}
}

@article{Duan20262,
  title = {A unified anisotropic {VDM--PFM} theory for failure in fiber-reinforced composite materials},
  volume = {215},
  url = {http://dx.doi.org/10.1016/j.jmps.2026.106739},
  DOI = {10.1016/j.jmps.2026.106739},
  journal = {Journal of the Mechanics and Physics of Solids},
  author = {Duan,  Ya and Yu,  Yuanfeng and Ren,  Huilong and Bie,  Yehui and Zhuang,  Xiaoying and Rabczuk,  Timon},
  year = {2026},
  pages = {106739}
}

@article{Yoshioka2021,
  title = {Variational phase-field fracture modeling with interfaces},
  volume = {384},
  url = {http://dx.doi.org/10.1016/j.cma.2021.113951},
  DOI = {10.1016/j.cma.2021.113951},
  journal = {Computer Methods in Applied Mechanics and Engineering},
  author = {Yoshioka,  Keita and Mollaali,  Mostafa and Kolditz,  Olaf},
  year = {2021},
  pages = {113951}
}

@article{Zhou2022,
  title = {An interface-width-insensitive cohesive phase-field model for fracture evolution in heterogeneous materials},
  volume = {256},
  url = {http://dx.doi.org/10.1016/j.ijsolstr.2022.111980},
  DOI = {10.1016/j.ijsolstr.2022.111980},
  journal = {International Journal of Solids and Structures},
  author = {Zhou,  Q.Q. and Wei,  Y.G. and Zhou,  Y.C. and Yang,  L.},
  year = {2022},
  pages = {111980}
}

@article{Bian2025,
  title = {Adaptive phase-field cohesive-zone model for simulation of mixed-mode interfacial and bulk fracture in heterogeneous materials with directional energy decomposition},
  volume = {443},
  url = {http://dx.doi.org/10.1016/j.cma.2025.118062},
  DOI = {10.1016/j.cma.2025.118062},
  journal = {Computer Methods in Applied Mechanics and Engineering},
  author = {Bian,  Pei-Liang and Liu,  Qinghui and Zhang,  Heng and Qing,  Hai and Schmauder,  Siegfried and Yu,  Tiantang},
  year = {2025},
  pages = {118062}
}

@article{Najmeddine2025,
  title = {Coupled large deformation phase-field and cohesive zone model for crack propagation in hard-soft multi-materials},
  volume = {196},
  url = {http://dx.doi.org/10.1016/j.jmps.2024.106016},
  DOI = {10.1016/j.jmps.2024.106016},
  journal = {Journal of the Mechanics and Physics of Solids},
  author = {Najmeddine,  Aimane and Gupta,  Shashank and Moini,  Reza},
  year = {2025},
  pages = {106016}
}

@article{Paggi2017,
  title = {Revisiting the problem of a crack impinging on an interface:A modeling framework for the interaction between the phase field approach for brittle fracture and the interface cohesive zone model},
  volume = {321},
  url = {http://dx.doi.org/10.1016/j.cma.2017.04.004},
  DOI = {10.1016/j.cma.2017.04.004},
  journal = {Computer Methods in Applied Mechanics and Engineering},
  author = {Paggi,  M. and Reinoso,  J.},
  year = {2017},
  pages = {145–172}
}

@article{Zhang2020,
  title = {Modelling distinct failure mechanisms in composite materials by a combined phase field method},
  volume = {232},
  url = {http://dx.doi.org/10.1016/j.compstruct.2019.111551},
  DOI = {10.1016/j.compstruct.2019.111551},
  journal = {Composite Structures},
  author = {Zhang,  Peng and Feng,  Yaoqi and Bui,  Tinh Quoc and Hu,  Xiaofei and Yao,  Weian},
  year = {2020},
  pages = {111551}
}

@article{deFrancqueville2020_2,
  title = {Comparison of the finite strain macroscopic behavior and local damage of a soft matrix highly reinforced by spherical or polyhedral particles},
  volume = {84},
  ISSN = {0997-7538},
  url = {http://dx.doi.org/10.1016/j.euromechsol.2020.104070},
  DOI = {10.1016/j.euromechsol.2020.104070},
  journal = {European Journal of Mechanics - A/Solids},
  publisher = {Elsevier BV},
  author = {de Francqueville,  Foucault and Gilormini,  Pierre and Diani,  Julie and Vandenbroucke,  Aude},
  year = {2020},
  pages = {104070}
}

@article{Weng2019,
  title = {Three-dimensional multi-particle {FE} model and effects of interface damage,  particle size and morphology on tensile behavior of particle reinforced composites},
  volume = {209},
  ISSN = {0263-8223},
  url = {http://dx.doi.org/10.1016/j.compstruct.2018.11.008},
  DOI = {10.1016/j.compstruct.2018.11.008},
  journal = {Composite Structures},
  publisher = {Elsevier BV},
  author = {Weng,  Lin and Fan,  Tongxiang and Wen,  Mao and Shen,  Yao},
  year = {2019},
  pages = {590–605}
}

@article{Yang2022,
  title = {Effect of particle morphology on mechanical behaviour of highly particle-filled composites},
  volume = {227},
  ISSN = {0020-7403},
  url = {http://dx.doi.org/10.1016/j.ijmecsci.2022.107446},
  DOI = {10.1016/j.ijmecsci.2022.107446},
  journal = {International Journal of Mechanical Sciences},
  publisher = {Elsevier BV},
  author = {Yang,  Zheng and Kang,  Ge and Liu,  Rui and Chen,  Pengwan},
  year = {2022},
  pages = {107446}
}

@article{NguyenThanh2022,
  title = {Multi phase-field modeling of anisotropic crack propagation in 3D fiber-reinforced composites based on an adaptive isogeometric meshfree collocation method},
  volume = {393},
  ISSN = {0045-7825},
  url = {http://dx.doi.org/10.1016/j.cma.2022.114794},
  DOI = {10.1016/j.cma.2022.114794},
  journal = {Computer Methods in Applied Mechanics and Engineering},
  publisher = {Elsevier BV},
  author = {Nguyen-Thanh,  Nhon and Li,  Weidong and Huang,  Jiazhao and Zhou,  Kun},
  year = {2022},
  pages = {114794}
}

@book{Zienkiewicz2000,
  title={The finite element method: solid mechanics},
  author={Zienkiewicz, Olgierd Cecil and Taylor, Robert Leroy},
  volume={2},
  year={2000},
  publisher={Butterworth-heinemann}
}

@article{Tang2019,
  title = {Phase field modeling of fracture in nonlinearly elastic solids via energy decomposition},
  volume = {347},
  ISSN = {0045-7825},
  url = {http://dx.doi.org/10.1016/j.cma.2018.12.035},
  DOI = {10.1016/j.cma.2018.12.035},
  journal = {Computer Methods in Applied Mechanics and Engineering},
  publisher = {Elsevier BV},
  author = {Tang,  Shan and Zhang,  Gang and Guo,  Tian Fu and Guo,  Xu and Liu,  Wing Kam},
  year = {2019},
  month = Apr,
  pages = {477–494}
}

@article{NajafiKoopas2025,
  title = {Comparative Analysis of Phase-Field and Intrinsic Cohesive Zone Models for Fracture Simulations in Multiphase Materials with Interfaces: Investigation of the Influence of the Microstructure on the Fracture Properties},
  volume = {15},
  ISSN = {2076-3417},
  url = {http://dx.doi.org/10.3390/app15010160},
  DOI = {10.3390/app15010160},
  number = {1},
  journal = {Applied Sciences},
  publisher = {MDPI AG},
  author = {Najafi Koopas,  Rasoul and Rezaei,  Shahed and Rauter,  Natalie and Ostwald,  Richard and Lammering,  Rolf},
  year = {2025},
  pages = {160}
}

\end{document}